\documentstyle[12pt,aaspp4,flushrt]{article}

\newbox\grsign \setbox\grsign=\hbox{$>$} \newdimen\grdimen \grdimen=\ht\grsign
\newbox\simlessbox \newbox\simgreatbox
\setbox\simgreatbox=\hbox{\raise.5ex\hbox{$>$}\llap
     {\lower.5ex\hbox{$\sim$}}}\ht1=\grdimen\dp1=0pt
\setbox\simlessbox=\hbox{\raise.5ex\hbox{$<$}\llap
     {\lower.5ex\hbox{$\sim$}}}\ht2=\grdimen\dp2=0pt
\def\simgreat{\mathrel{\copy\simgreatbox}}
\def\simless{\mathrel{\copy\simlessbox}}

\def\vol#1  {{{#1}{\rm,}\ }}

\def\aj{{AJ}, }  
\def\apj{{ApJ}, } 
\def\apjs{{ApJS}, } 
\def\pasp{{PASP}, }  
\def\mnras{{MNRAS}, } 
\def\aa{{A\&A}, }     
\def\aasup{{A\&AS}, } 
\def\etal{{\it et al.}\ }

\def\clalpha{{CL1324+3011}}
\def\cldelta{{CL1604+4304}}
\def\cleps{{CL1604+4321}}
\def\kms{km s$^{-1}$}
\def\ie{{\it i.e.,\ }}
\def\eg{{\it e.g.,\ }}
\def\ih65{$h_{65}^{-1}$\ }
\def\sh65{$h_{65}^{-2}$\ }
\def\ch65{$h_{65}^{3}$\ }
\def\h65{$h_{65}$\ }

\slugcomment{Accepted for Publication in {\it The Astronomical Journal}}

\begin{document}
\title{A Study of Nine High-Redshift Clusters of Galaxies: IV. Photometry and Spectra of Clusters 1324+3011 and 1604+4321}
\author{Marc Postman}
\affil{Space Telescope Science Institute\altaffilmark{1}, 3700 San Martin Drive, Baltimore, MD 
21218}
\affil{Electronic mail: postman@stsci.edu}
\author{Lori M. Lubin\altaffilmark{2,3}}
\affil{Palomar Observatory, California Institute of Technology,  Pasadena, 
CA 91125}
\affil{Electronic mail: lml@pha.jhu.edu}
\author{J. B. Oke}
\affil{Palomar Observatory, California Institute of Technology, Pasadena, 
CA 91125}
\affil{and}
\affil{Dominion Astrophysical Observatory, 5071 W. Saanich Road, Victoria, 
BC V9E 2E7}
\affil{Electronic mail: Bev.Oke@nrc.ca}

\altaffiltext{1}{Space Telescope Science Institute is operated by the
Association of Universities for Research in Astronomy, Inc.,
under contract to the National Aeronautics and Space Administration.}
\altaffiltext{2}{Hubble Fellow}
\altaffiltext{3}{Current address : Department of Physics and
Astronomy, Johns Hopkins University, Baltimore, MD 21218}

\baselineskip 14pt

\clearpage

\begin{abstract}

New photometric and spectroscopic observations of galaxies in the directions
of three distant clusters are presented as part of our on-going
high-redshift cluster survey. The clusters are CL1324+3011 at $z =
0.76$, CL1604+4304 at $z = 0.90$, and CL1604+4321 at $z = 0.92$.
We have spectroscopically confirmed cluster membership for 20 to 40 galaxies
in each system and have also obtained spectra for over 280 field galaxies
spanning the range $0 < z < 2.5$.
Kinematic estimates of the mass within the central 770\ih65 kpc of each
cluster are in excess of $8 \times 10^{14}$ \ih65 M$_{\odot}$. 
The observed x-ray luminosities in these clusters 
are at least a factor of 3 smaller
than those observed in clusters with similar 
velocity dispersions at $z \le 0.4$. 

These clusters contain a significant population of elliptical-like galaxies, 
although these galaxies are not nearly as dominant as in massive clusters at
$z\le 0.5$.  We also find a large population of blue cluster members.
Defining an active galaxy as one in which the rest equivalent width of
[OII] is greater than 15\AA, the fraction of active cluster galaxies,
within the central 1.0 \ih65 Mpc, is 45\%. In the field population, we
find that 65\% of the galaxies with redshifts between $z = 0.40$ and
$z = 0.85$ are active, while the fraction is 79\% for field galaxies at $z>0.85$. 
The star formation rate normalized by the rest AB $B-$band magnitude,
SFRN, increases as the redshift increases at a given evolving luminosity. 
At a given redshift, however, SFRN decreases linearly with increasing luminosity
indicating a remarkable insensitivity of the star formation rate to
the intrinsic luminosity of the galaxy over the range $-18 \ge {\rm ABB} \ge -22$. 
Cluster galaxies in the central 1\ih65 Mpc regions exhibit depressed
star formation rates and contain a larger fraction of galaxies with {\it ``k"} type
spectra. The star formation rates in galaxies lying between $1 - 2.5$\ih65 Mpc from the
cluster centers, however, are in good agreement with that in galaxies 
in the general field at similar redshifts.
The spectroscopic and photometric properties of the
cluster galaxies are well fit by Bruzual-Charlot solar metallicity,
constant-age (4.8 Gyr at $z = 0.9$), variable tau models.
Metallicities in these clusters must be at least 0.2 of solar, and a
significant amount of dust extinction is unlikely.  

We are able to measure significant evolution in the $B$-band luminosity function
over the range $0.1 \le z \le 1$. The characteristic luminosity increases
by a factor of 3 with increasing redshift over this range.
This result is consistent with an analysis of the luminosities of
the brightest cluster galaxies in these clusters. The BCGs are
typically twice as luminous as their current epoch counterparts. 

\end{abstract}


\clearpage

\section{Introduction}
\label{sec:intro}

Our understanding of the cosmic history of galaxy clusters is slowly
maturing due to an ever growing series of observations including faint
spectroscopic data (especially those obtained at the Keck and VLT
observatories), deep optical and near-infrared (NIR) imaging from the ground and in
space, morphological data from the Hubble Space Telescope (HST), 
and constraints on the evolution
of the intracluster medium (ICM) from ROSAT, ASCA, and XMM.  When such
observations are applied to complete, objectively derived catalogs of
clusters, the constraints placed on cluster formation and evolution
scenarios can become quite confined.  Cluster evolution is inherently
complex both because clusters are not closed systems and because the 3
main mass components (dark matter, ICM, and galaxies) evolve
differently.  As a consequence, different cluster parameters evolve on
different timescales depending on the thermal and dissipative
properties of the mass component(s) which most strongly control each
cluster parameter.

The properties of clusters at redshifts as low as $z \sim 0.4$
(lookback times of $\sim 0.33$ the present age of the universe)
already exhibit significant departures from their current epoch
counterparts.  The broad-band color distribution of the early-type
galaxy population show significant bluing; the observed trend is
consistent with passive stellar evolution and a relatively well
synchronized initial starburst epochs occurring at $z > 2.5$ (\eg
Arag\'on-Salamanca \etal 1993; Stanford, Eisenhardt, \& Dickinson
1995; Ellis \etal 1997). In contrast, the relative abundance and the
spectral characteristics of the disk galaxies in clusters appear to have
evolved significantly over the last third of a Hubble time (Dressler
\etal 1997; Poggianti \etal 1999). Indeed, the infall and processing
of disk galaxies in clusters appears to continue right up to the
present day (Adami et al.\ 1998).  However, even at $z \sim 0.4$ the
properties of cluster galaxies are noticeably different from those in
the surrounding field.

At redshifts of $z \simgreat 0.8$, the study of massive clusters
provides particularly important constraints on the physical processes
that dominate the formation of their member galaxies (and on
cosmological parameters) because the amplitude of evolutionary effects
and the differences between competing theories are quite large (\eg
Bower, Kodama, \& Terlevich 1998).  The observations of clusters at $z
\sim 1$ that exist today are limited to a handful of clusters, largely
because so few systems are known. While new distant cluster surveys
will remedy this lack of targets (\eg Gladders 2000; Postman \etal
2001; Gonzales \etal 2001), several intriguing observations of
existing systems already suggest that the $z \sim 1$ epoch is one at
which clusters and their member galaxies exhibit significant
differences from their descendants at $z \simless 0.5$. For example,
there is evidence that a significant fraction of the early type
galaxies in MS-1054 ($z = 0.83$) are merging -- something that is not
seen in the vast majority of current epoch cluster ellipticals -- and
which results in a significant increase in the scatter in their color
-- magnitude relation relative to what is observed at $z < 0.5$ (van
Dokkum \etal 2000). The fraction of cluster galaxies with evidence for
active star formation is also significantly higher at $z > 0.75$
($\sim 50$\%) than the fraction at the current epoch or even $z
\sim 0.5$ (Postman, Lubin, \& Oke 1998). At the same time, the global
properties of the ICM seem well established even by $z = 0.8$ as there
appears to be little evolution observed in the $L_x - T_x$ relation
out to these redshifts (Mushotzky \& Scharf 1997; Donahue \etal 1998),
although the fraction of clusters with asymmetric x-ray gas and asymmetric
galaxy surface density distributions increases noticeably (\eg Lubin \& Postman
1996; Gioia \etal 1999).

It is quite important, therefore, to conduct statistically complete
spectroscopic and photometric surveys of many $z \sim 1$ clusters,
covering as broad a range in global cluster properties as possible, in
order to understand the breadth of the physics associated with cluster
galaxy formation and evolution and, in particular, how the cluster
environment modifies the path of galaxy evolution. Focusing solely on
one component of the cluster population (\eg ellipticals) will only
reveal part of the story and may even result in a biased
interpretation of the timescales for galaxy formation (\eg van Dokkum
\& Franx 2001).  In 1995, we began an extensive spectroscopic and
imaging program to study nine candidate clusters of galaxies at $z
\simgreat 0.7$ (Oke, Postman \& Lubin 1998; hereafter Paper I) in an
attempt to establish an evolutionary reference sample of clusters
analogous to the MORPHs survey performed at $z \sim 0.5$ (Smail \etal
1997).  Our sole spectroscopic target selection criterion is the
galaxy's $R$-band magnitude.  No color selection is applied in order
to assure that a broad range of cluster galaxy types are included in
the survey. The clusters themselves span a relatively broad range in
x-ray luminosity and richness.

In this paper, we present our measurements and interpretations of the
kinematic and spectrophotometric properties of the galaxies in the
clusters \clalpha\ ($z = 0.757$), \cldelta\ ($z = 0.897$), and \cleps\
($z = 0.924$) -- the three most massive systems in our survey.
\cleps\ is of particular interest because its comoving spatial
separation from the better known cluster \cldelta\ is about 15 Mpc and
their cosmologically corrected radial velocity separation is 4350
\kms. Lubin \etal (2000) provide extensive observational evidence that
suggests \cldelta\ and \cleps\ are indeed two members of a rich
supercluster. Some of the spectrophotometric and kinematic results for
\cldelta\ have already been published (Postman, Lubin, \& Oke 1998;
hereafter Paper II) along with a study of the morphological properties
of its galaxy population (Lubin \etal 1998; hereafter Paper III).
Here we provide both new and improved constraints on the cluster mass
estimates, on the evolution of the galaxy luminosity function, and on
the spectral characteristics, star formation rates, and stellar
population ages of galaxies for all three of the above clusters in
this paper.  A separate paper (Lubin \etal 2001) will present results
on the morphological make-up of the cluster galaxies and on the
relationship between galaxy morphology and local density in these
clusters.

A brief summary of the observations is presented in
\S\ref{sec:obs}. Our measurements of the cluster masses and
mass-to-light ratios are presented in \S\ref{sec:mtl} and our
constraints on the evolution of the galaxy luminosity function (using
both our cluster and field samples) are included in
\S\ref{sec:lfevol}.  Comparisons between spectral synthesis models and
the observed spectrophotometric data are described in
\S\ref{sec:modcompare}.  A discussion of the characteristics of the
brightest cluster galaxies, their evolution, and the ``red sequence"
is given in \S\ref{sec:clusprop}.  Analyses of the star formation
rates, stellar population ages, and an improved estimate of the
fraction of active galaxies in these clusters are contained in
\S\ref{sec:emlsfr} and the correlation of these properties with
clustocentric radius is described in \S\ref{sec:clstruc}. A discussion
of the above results along with a summary of our conclusions are given
in \S\ref{sec:conclude}.  We adopt $h_{65} \equiv {\rm H}_o / (65$ km
s$^{-1}$ Mpc$^{-1}) = 1$, $\Omega_o = 0.2$, and $\Lambda = 0$
throughout this paper.

\section{Observations}
\label{sec:obs}

The photometric and spectroscopic 
observations of \clalpha, \cldelta, and \cleps\ were conducted using 
the Low Resolution Imaging Spectrograph at the W. M. Keck Observatory
(Oke \etal 1995).
We provide a brief description of the observations below but refer
the reader to Paper I for the full details and to Paper II for the
specifics on observations of \cldelta.

\subsection{Broadband Imaging }
\label{sec:img}

Broadband $BVRI$ images of \clalpha\ and \cleps\ were acquired during
a series of observing runs spread over a 3 year period (March 1994
through August 1997) and were obtained under photometric conditions.
We use the SExtractor (Bertin \& Arnouts 1996) source detection and
classification package to generate object positions, classifications,
and aperture photometry from the reduced images. This differs from our
work in Paper II where we used the FOCAS (Valdes 1982) package.  The
aperture photometry from SExtractor is more accurate than that from
FOCAS because SExtractor employs pixel masks to exclude light from
sources surrounding the object being photometered. We have also
reprocessed the image data from \cldelta\ through the SExtractor
package to enable consistent comparisons with the clusters \clalpha\
and \cleps.  The SExtractor detection parameters were set to match the
FOCAS parameters described in Paper I.  The FWHM of the PSF in the
images is $\sim0.8'',\ \sim1.0'',\ {\rm and}\ \sim0.9''$ for \clalpha,
\cldelta, and \cleps, respectively.  A fixed aperture diameter of
6.0$''$ was used to compute the aperture photometry. This diameter
corresponds to projected metric radii of 21.7 \ih65 kpc and 22.9\ih65
kpc at $z = 0.76$ and $z = 0.90$, respectively.  This aperture choice
is consistent with that used for faint galaxy photometry in other
intermediate redshift cluster analyses ({\it e.g.}, Arag\'on-Salamanca
\etal 1993, Smail, Ellis \& Fitchett 1994, Barger \etal 1996).

The number of objects detected and measured is 4952, 3905, and 6322 in
\clalpha, \cldelta, and \cleps, respectively.  Not all objects are
detected in all four passbands partly because, in some cases, small
positional offsets exist between the various filter images whenever
the data were obtained on separate observing runs. The number of
objects detected refers to only those objects that are detected in
{\it at least} two of the four bands. Spectroscopic target
selection, however, is based solely on an object's $R-$band magnitude (see
\S\ref{sec:spec} below). We also note that all spectroscopically confirmed
cluster members are detected in at least 3 of the four passbands so
the choice to catalog objects in the above way does not hinder our study
of the cluster galaxy populations. Each object is assigned a Keck
identification number (based on its location in a declination-ordered
catalog) for convenient referencing.  There is a very bright star (V
$\approx$ 10.3) in the southern part of the \cleps\ and objects within
$\sim30''$ of the star are not photometrically analyzed or classified.

All $BVRI$ magnitudes have been converted to absolute AB magnitudes
using the relations given in Paper I.  Models show that the relation
between AB magnitude and log $\nu$, where $\nu$ is the observed
effective frequency of each broadband filter, is nearly linear with a
small curvature that changes sign when proceeding from hot to cool
stellar populations.  We have, thus, chosen to characterize the
observed broadband energy distributions by the slope, hereafter
referred to as the slope $b$, of a linear least squares fit to the
measured AB magnitude as a function of log $\nu$.  The slope $b$
provides a more robust indicator of the overall broadband SED than any
individual color measure such as $B-V$, $V-R$, or $R-I$.

\begin{figure}
\plotone{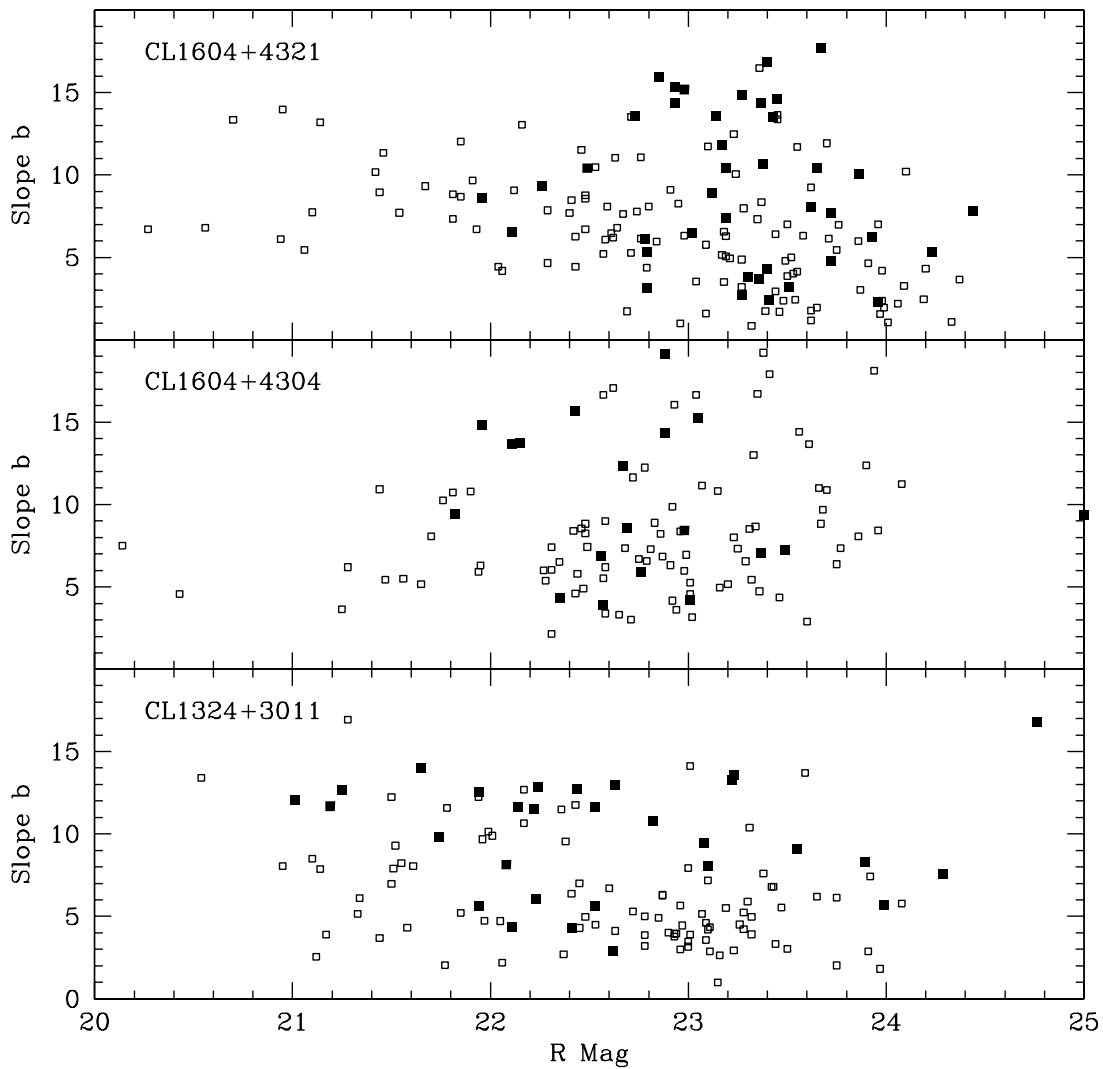}
\caption{The slope of the broadband spectral energy distribution, $b$,
as a function of $R$ magnitude for the fields centered
on the clusters \clalpha, \cldelta, and \cleps.
Solid squares are spectroscopically confirmed
cluster members while open squares are non-members.
The slope $b$ is strongly correlated with the usual broadband
color measurements. For example, $(V-R) \approx b/13$. }
\label{figCC}
\end{figure}

The slope $b$ of the cluster and field galaxies are shown as a
function of $R$ magnitude in Figure~\ref{figCC} for the \clalpha,
\cldelta, and \cleps\ fields.  Spectroscopically confirmed cluster
members are indicated by the filled symbols. The distributions in
slope $b$ are very similar for both cluster and field galaxies; there
is only a hint of the red color ridge normally seen in similar plots
of clusters at redshifts near 0.5. The lack of a strong red envelope
is a consequence of the rest wavelengths of the optical bands used and
the relatively young age of the cluster galaxies.  A more detailed
discussion of this result is provided in \S\ref{sec:redenv}.

\subsection{Spectroscopic Observations}
\label{sec:spec}

\begin{figure}
\plotone{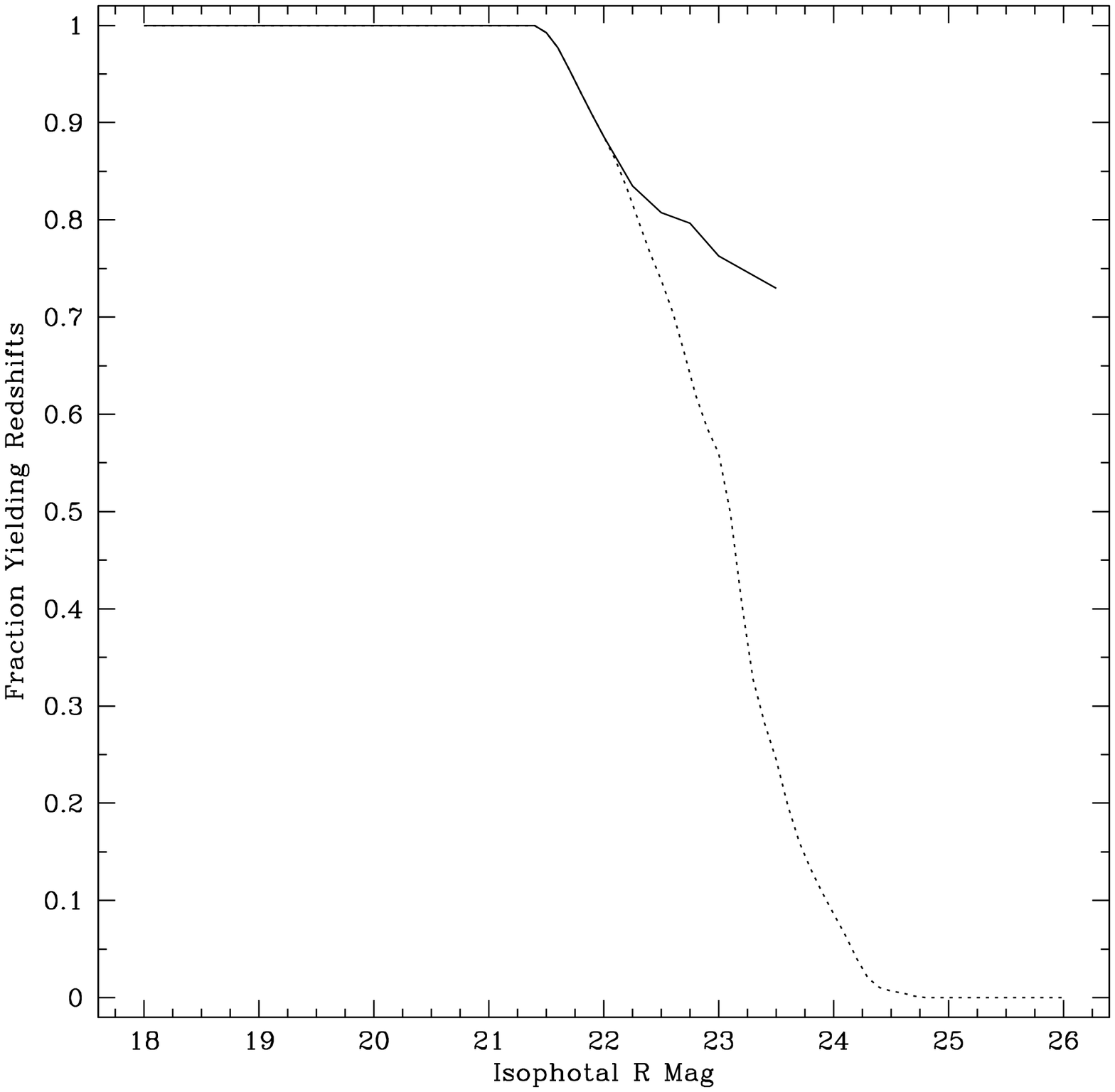}
\caption{The solid curve shows the fraction of spectra for which redshifts were
successfully measured as a function of $R$ magnitude. The dashed curve shows
the fraction of all galaxies in the field with measured redshifts as a
function of $R$ magnitude.}
\label{figzcomp}
\end{figure}

Spectra were obtained using several slit masks (7 for \clalpha, 6 for
\cleps) for galaxies in the range $18 \simless {\rm R} \simless 23.5$
(the precise magnitude limits are 23.45 for \clalpha\ and 23.60 for
\cleps). The bright limit was imposed to avoid saturating the
detector. The galaxies observed spectroscopically are distributed over
a $\sim2' \times \sim7'$ region centered on each cluster. 
Using the procedures described in Paper I, we succeeded in
obtaining redshifts for 88\% of the objects observed.  Table 1
summarizes the yields and Figure~\ref{figzcomp} shows our redshift
success rate as a function of $R$ magnitude.  Failure to measure a
redshift is mostly due to insufficient signal-to-noise ratio, but
occasionally we do encounter good S/N spectra with no identifiable
features. Some of the latter spectra may be very low redshift objects
without significant absorption or emission features redwards of
5000\AA\ in the rest frame.  In \clalpha\ there is one QSO or AGN,
Keck \#3933, with a redshift of 1.0750.  In \cleps, Keck \#1339 is the
sole QSO at a redshift of 2.4970.

Tables 2, 3, and 4 provide the key photometric and spectroscopic
parameters for all galaxies with measured redshifts for the survey
fields centered on \clalpha, \cldelta, and \cleps.  The Keck object
identification number is given in column 1 and if the number is
preceeded by an asterisk then the galaxy is considered to be a member
of the cluster (see \S\ref{sec:mtl} for details). The brightest
cluster galaxy in each case is denoted by a double asterisk. We note
that the identification numbers used in Paper II for \cldelta\ differ
from those in Table 3 because we used different object detection
software.  The absolute AB magnitudes in our 4 passbands, $ABB,\ ABV,\
ABR,\ {\rm and}\ ABI$, are given in columns 2 through 5. Geocentric
redshifts and our measure of the redshift quality (see Paper I) are
given in columns 6 and 7. The slope $b$ (defined above in units of AB
mag per unit interval of log $\nu$) is listed in column 8. The
contents of the remaining columns are described in subsequent
sections.

\section{Cluster Mass and M/L Estimates}
\label{sec:mtl}

As in Paper II, we derive cluster masses based on three popular virial
theorem mass estimators: the traditional pairwise mass estimator,
M$_{PW}$, the projected mass estimator, M$_{PM}$ (Bahcall \& Tremaine
1981; Heisler, Tremaine, \& Bahcall 1985), and the ring-wise mass
estimator, M$_{RW}$ (Carlberg \etal 1996). The mathematical
definitions of these estimators are given in equations 2 -- 6 in Paper
II.  Each estimator has its strengths and weaknesses. The M$_{PW}$
estimate does not require one to specify a cluster center. However
M$_{PM}$ and M$_{RW}$, which do require a center to be defined, tend
to be much more robust against interlopers.  The radial cluster
velocity dispersion, a necessary parameter in virial mass estimation,
is accurately determined from the redshifts for $22-41$ cluster
members in each system.  Velocity dispersions are computed by first
defining a broad redshift range, typically $\Delta z = \pm 0.06$, in
which to conduct the calculations.  This range is manually chosen to
be centered on the approximate redshift of the cluster.  We then
compute the bi-weight mean and dispersion of the velocity distribution
(Beers, Flynn, \& Gebhardt 1990) and identify the galaxy with the
largest deviation from the mean. Velocity offsets from the mean are
taken to be $\Delta v = c(z - \overline{z})/(1 + \overline{z})$ which
corrects for cosmological and relativistic effects. In the case of
bi-weight statistics, $\overline{z}$ is the median of the
distribution.  If the galaxy with the largest velocity deviation
differs from the bi-weight median by either more than 3$\sigma$ or by
more than 3500 km s$^{-1}$, it is excluded, and the computations are
redone. The procedure continues until no further galaxies satisfy the
above criteria.  The 3500 km s$^{-1}$ limit is based on extensive data
available for low $z$ clusters. For example, 95\% of the galaxies
within the central 4.6\ih65 Mpc region of the Coma cluster and with
$cz \le 12,000$ km s$^{-1}$ lie within $\pm 3500$ km s$^{-1}$ of the
mean Coma redshift. This clipping procedure is conservative and does
not impose a Gaussian distribution on the final redshift distribution
(see {\it e.g.,} CL0023+0423 in Paper II).

\begin{figure}
\plotone{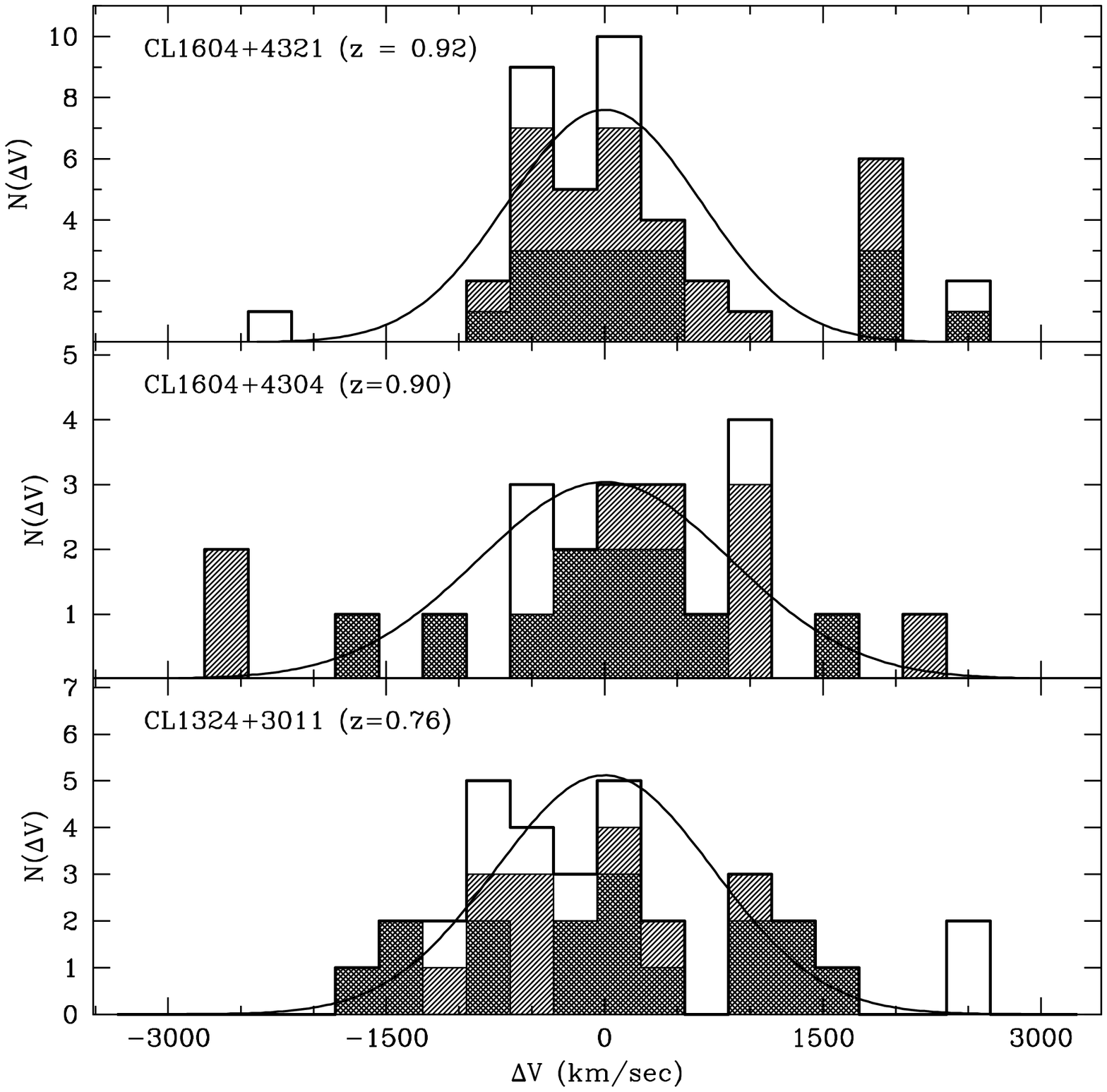}
\caption{Histogram of the relativistically corrected velocity offsets
for \clalpha, \cldelta, and \cleps. Offsets are relative
to the mean cluster redshift. Best fit Gaussian distributions are shown
for comparison. The darkest histograms include only those galaxies within
the central 385\ih65 kpc. The intermediate shading represents the galaxies
within the central 770\ih65 kpc. The unshaded histograms show the distributions
for all available data. }
\label{figvdisp}
\end{figure}

Figure~\ref{figvdisp} shows histograms of the velocity offsets
relative to the mean cluster redshifts for the three clusters. The
derived kinematic parameters, including the mean $z$, dispersion, and
mass estimates, are provided in Table 5. For each cluster, we give the
results using all available redshift data (no radius limit), as well
as the results for those galaxies within the central 385 and 770\ih65
kpc regions. The radially limited results are used to derive central
mass-to-light (M/L) ratios.  The results for \cldelta\ were originally
published in Paper II, but we include them here for convenience.  The
derived masses within the central 770\ih65 kpc regions of these 3
clusters are all in excess of $8 \times 10^{14}$\ih65 M$_{\odot}$ (for
the M$_{PW}$ estimator) and the projected and ring-wise mass
estimators yield central values $\simgreat 10^{15}$\ih65 M$_{\odot}$.
Each of the clusters has also been detected in X-rays by ROSAT
(Castander \etal 1994), providing further evidence that these clusters
have developed deep potential wells.  The $0.1 - 2.4$ keV X-ray
luminosities are $L_x = (4.80\pm1.49) \times 10^{43}$\sh65 erg
s$^{-1}$, $L_x = (6.39\pm1.37) \times 10^{43}$\sh65 erg s$^{-1}$, and
$L_x \le $ 4.09 $\times 10^{43}$\sh65 erg s$^{-1}$ (3$\sigma$ upper
limit) for \clalpha, \cldelta, and \cleps, respectively.  However,
based on the local $L_x - \sigma$ relation (Edge \& Stewart 1991),
these x-ray luminosities are low for the derived velocity dispersions.
Similar trends have been seen in other studies of intermediate
redshift ($0.4 < z < 0.7$) clusters (Couch \etal 1991; Holden \etal
1997; Gioia \etal 1999) suggesting that these systems are dynamically
young and, consequently, the relationship between the temperature of
their x-ray emitting ICM and their total gravitational mass is still
undergoing significant evolution.

The central $BVR$ mass-to-light ratios, in solar units, for these
clusters are given in Table 6 using the projected mass estimate,
$M_{PM}$, from Table 5.  The errors shown include the formal
uncertainties in both the mass and luminosity estimates but do not
include any systematic error estimates.  The cluster luminosities are
computed as described in \S3.2 of Paper II.  The M/L values in Table 6
are based on a non-evolving cluster luminosity function. If we assume
that $M^*$ evolves as $M^*(z) \approx M^*(0) - z$ (Lilly \etal 1995;
see also \S\ref{sec:lfevol}), the M/L ratios increase by about 10\%,
relative to the non-evolving calculation. The dependence on the
evolution of $M^*$ is through the correction applied to the total
luminosity for the assumed missing faint end of the cluster luminosity
function (see equation 8 in Paper II).  Systematic errors in the M/L
ratios can be significant, however, because virial equilibrium may not
be fully achieved [Small \etal (1998) have shown worst case
overestimates in the virial mass by a factor of 2 can occur for
marginally bound systems] and because the background subtraction is
performed statistically, which can be problematic at high redshifts as
the light from foreground sources dominates. We minimize this latter
effect by confining the luminosity computation to galaxies with
apparent magnitudes spanning the range defined by the
spectroscopically confirmed members. A correction must still be
applied to this luminosity sum, however, to reflect an integration to
a common fiducial absolute luminosity, taken to be $-11.15$ +
5log$_{10}$\h65 in this case.

The central M/L ratios for these three distant clusters are
consistent, in mean and scatter, with those seen in their local
counterparts (\eg Girardi \etal 2000) covering a range of $100 \simless M/L_B
< 350$\h65, a range inconsistent with an $\Omega_m = 1$ cosmology
($M/L_B = 2355$).  However, significant evolution in the M/L ratio on
cluster scales is easily masked by the above systematic errors.  A
more robust measurement of the M/L ratios for these 3 clusters
requires substantially more spectroscopic and morphological data than
is presently available.

\section{Luminosity Function Evolution}
\label{sec:lfevol}

The 321 galaxies in these 3 fields with redshifts in the range $0.1
\le z \le 1$ allow us to accurately constrain the evolution of the
rest AB $B$-band galaxy luminosity function over the last $\sim0.6$ of
a Hubble time. The $B$-band is chosen, as in Paper II, to eliminate or
minimize any extrapolation in the calculation of a rest-frame absolute
luminosity. We relate apparent and absolute magnitude using the
formalism of Equations 6, 9, and 10 in Gunn \& Oke (1975).  This
relation becomes :

\begin{eqnarray}
M_{AB_{\nu(1+z)}} = m_{AB_\nu} - 2.5~{\rm log}~\left[ \frac{9.00 \times
10^{20} {\pounds}^{2}_{q}(z)(1+z)} {H_o^2}\right]
\end{eqnarray}

\noindent where ${\pounds_q(z)}$ is given in Equation 9 of Gunn \& Oke
(1975). The best-fit stellar evolution model to the observed $BVRI$ AB
magnitudes (see \S\ref{sec:modcompare}) is then used to calculate the 
absolute magnitude at the rest $B$ wavelength from the above expression. 
The absolute magnitude in this
band is hereafter referred to as $M_{ABB}$.  For redshifts of $z <
0.92$ the redshifted $B$ filter position is within the observed
wavelength range, and an interpolation of the best-fit evolutionary
model can be made.  Above $z = 0.92$ the rest-frame $B$ filter
wavelength is above the observed $I$ band, and an extrapolation is
necessary.  This is done using the best-fit evolutionary model and
extrapolating to the appropriate frequency.  However, 90\% of the
galaxies in our $0.1 \le z \le 1$ sample lie below $z = 0.92$.  The
uncertainty in the resulting absolute AB can be estimated from the
uncertainty in the fit of the observations to the model.

\begin{figure}
\plotone{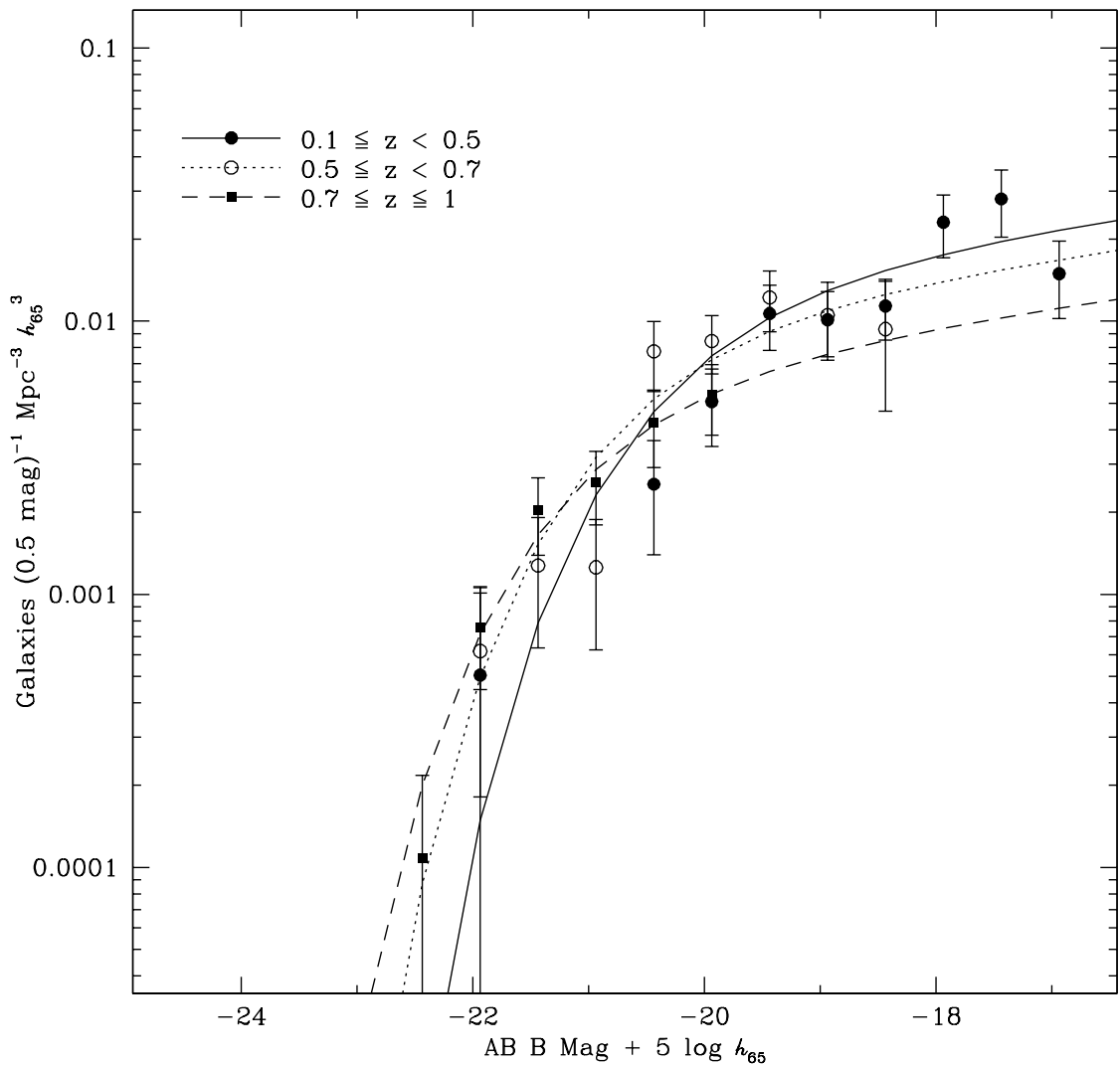}
\caption{The observed data and best fit Schechter
luminosity functions for field galaxies
in the range $0.1 \le z \le 1$. The best fits are constrained to have
faint end slopes of -1.15. See text and Table 7 for details.}
\label{figlffit}
\end{figure}

The luminosity function is computed in three redshift bins: $0.1 \le z
< 0.5$, $0.5 \le z < 0.7$, and $0.7 \le z \le 1.0$. Because the latter bin
contains the cluster members, we explicitly compute the field and cluster
luminosity functions separately for the $0.7 \le z \le 1.0$ interval. 
Each galaxy is weighted by the inverse
of $V/V_{max}$ to account for Malmquist bias and by the inverse of the
selection function (shown as the dashed curve in
Figure~\ref{figzcomp}) to account for the objects for which no
redshift was measured. The latter correction assumes that the
unmeasured galaxies have the same redshift distribution as a function
of apparent $R$ magnitude as those that do have measured redshifts. A
maximum likelihood method is then used to find the best fit Schechter
form luminosity function (Schechter 1976). Because our redshift survey
extends only 1.5 magnitudes fainter than the characteristic magnitude,
$M^*_{ABB}$, we constrain the faint end slope of the luminosity
function to have a value of $\alpha = -1.15$ (Marzke \etal
1998). Figure~\ref{figlffit} shows the observations and the best fits
for the field galaxy luminosity functions in the three redshift bins. 
The best-fit $M^*_{ABB}$ values (\h65 = 1) and other
relevant details about each redshift bin are provided in Table 7.

\begin{figure}
\plotone{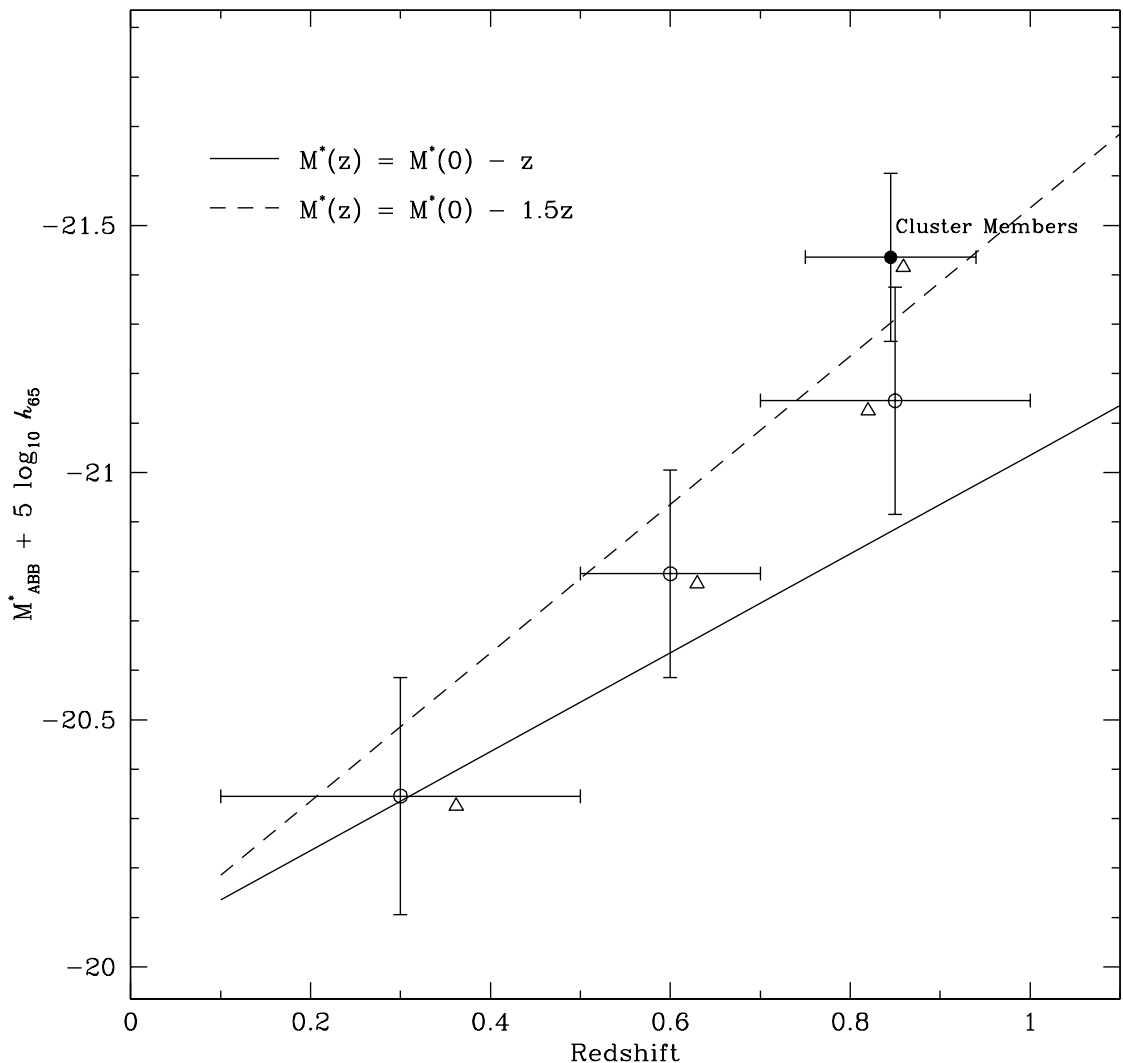}
\caption{The evolution of the luminosity function as traced
by the dependence of the characteristic magnitude (for
$\alpha = -1.15$) on redshift. Horizontal error bars
are set by the width of the redshift bins.
Open circles are the results for field galaxies, the filled circle
is the result for the cluster galaxies in the range $0.75 \le z \le 0.93$.
The open triangles show the mean redshift value in each bin.
The straight lines show a simple evolutionary model, parameterized as
$M^*(z) = M^*(0) - \beta z$, with $\beta = 1$ and $\beta = 1.5$.}
\label{figlfevol}
\end{figure}

The evolution of the $B$-band luminosity function is shown in
Figure~\ref{figlfevol}. In this figure we plot the best fit values of
$M^*_{ABB}$ as a function of redshift. The horizontal errors are just
set by the width of each bin in redshift space. Two simple models for
the evolution of the characteristic magnitude are shown.  Our data are
consistent with an evolution of the form $M^*(z) = M^*(0) - \beta z$
where $1 < \beta < 1.5$.  Such evolution is consistent with that
reported for blue field galaxies over the range $0.5 < z < 1$ (Lilly
\etal 1995). The $M^*_{ABB}$ derived for the cluster members is 
0.3 mag brighter than that for field galaxies in the range $0.7 \le z \le 1.0$. 
This is expected because the cluster sample includes many luminous early type
galaxies which are rarely found in low density environments.
The cluster $M^*_{ABB}$ at $z \sim 0.8$ is approximately 0.7 mag brighter
than that derived for clusters at $z < 0.2$ (\eg Colless 1989; Lumsden \etal 1997;
Valotto \etal 1997; Rauzy \etal 1998).

\section{Spectral Features and Comparisons with Spectral Synthesis Models}
\label{sec:modcompare}

\subsection{Models}

In order to derive constraints on the ``ages" of the galaxies in our
survey, we use the 1996 stellar evolution models of Bruzual \& Charlot
(hereafter BC96; see Bruzual \& Charlot 1993) because absolute energy
distributions can be generated over a broad wavelength range,
extending far below the Lyman limit, with a spectral resolution of
20\AA, only a factor 2 lower than our observed spectra (see Bruzual \&
Charlot 1993 for details).  The BC96 models can have different metal
abundances and we have used those with metallicities of $Z = 0.0200$
(solar) and $Z = 0.0040$ (0.2 solar). The simplest models are ssp
models where there is a single instantaneous burst of star formation
at time $t=0$.  One can also easily generate models where the star
formation rate falls exponentially with time.  We have used such
models with time constants ranging from 0.2 to 20.0 Gyr.  We will
refer to these as tau0.2, etc.  Unless otherwise noted, models used
have solar metallicity. Models with low metal abundance are listed as
tau0.6(0040), for example.

Because the models produce absolute energy distributions, it is
possible to duplicate almost any measurement that is performed on the
observational data.  The exception is the measurement of emission line
intensities where additional assumptions and calculations, discussed
in \S\ref{sec:emeqw}, are required.

\begin{figure}
\plotone{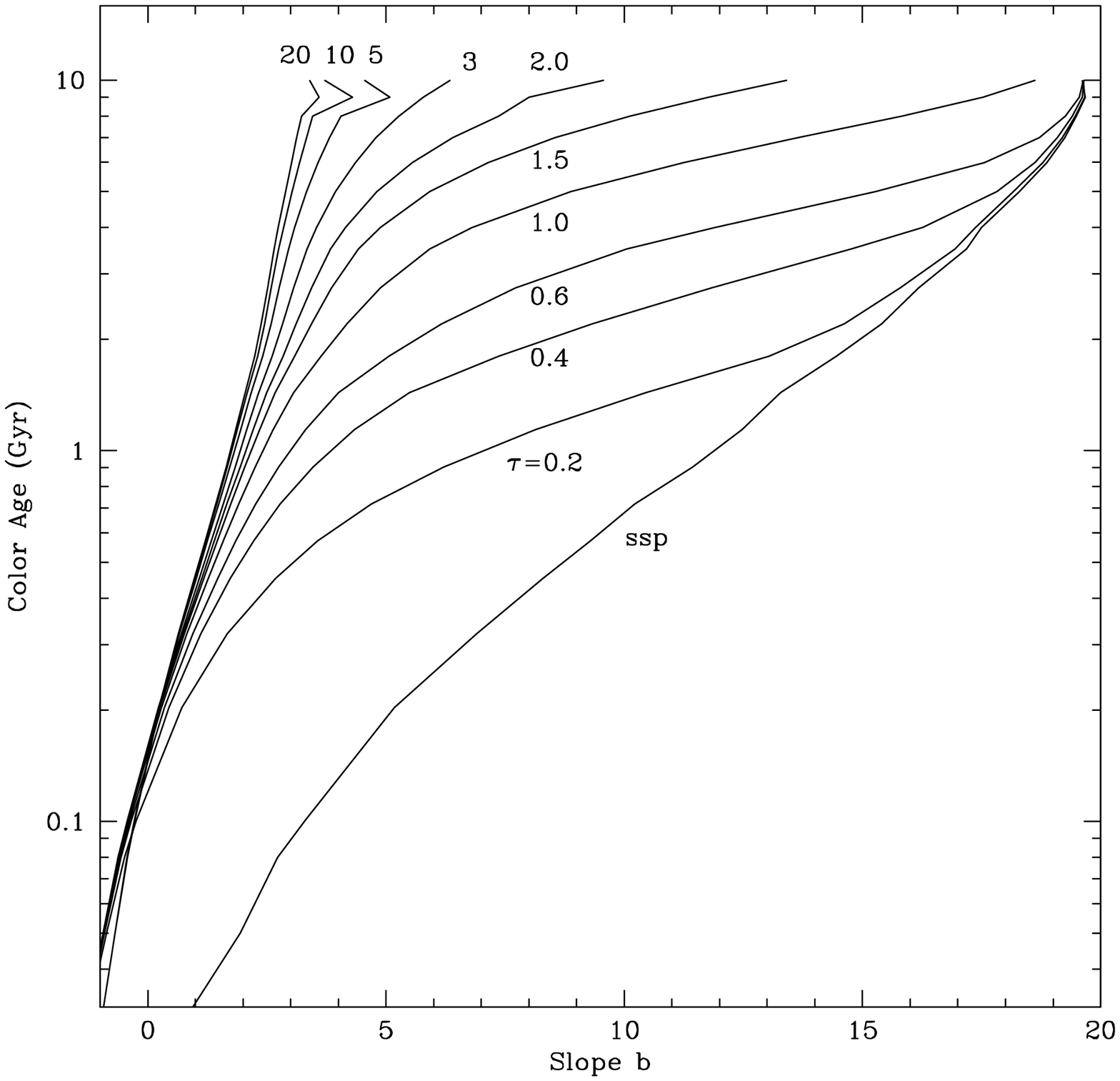}
\caption{The relation between the logarithm of the model age in Gyr and
the slope $b$.  The curves from bottom to top represent ssp, tau0.2, tau0.4,
tau0.6, tau1.0, tau 1.5, tau2.0, tau3.0, tau5.0, tau10.0, and tau20.0 models.}
\label{figFF}
\end{figure}

Any specific model is characterized by the details of the star
formation rate (SFR) as a function of the time and the age. A second
model parameter is the mean chemical composition. Clearly, there are a
large number of possible models, even without varying the chemical
composition, and many of them could fit a set of observations for a
single galaxy given the measurement uncertainties. We, therefore,
consider only two classes of models.  For ssp models there is only one
parameter and that is the age (in Gyr) after the initial burst of star
formation at time $t=0$.  For the tau models there are two parameters,
the $e$-folding time for decay of the SFR after $t=0$ and the age
after $t=0$. As described in Paper II, any model can be used to
generate the expected $BVRI$ AB magnitudes once the redshift is
specified.  The most striking characteristic of the models is the
change in the slope $b$, defined in \S\ref{sec:img}, as a function of
the age. For example, Figure~\ref{figFF} shows the slope $b$ as a
function of the logarithm of the model age in Gyr for the ssp model
and a series of tau models with different time constants, all computed
for a redshift of 0.90.

Over most of the range in $b$ the age derived from the absolute energy
distribution depends strongly on the choice of model.  At large values
of $b$ ($b \ge 15$), the models are very similar and essentially
independent of the decay time as long as the decay time constant is
less than 0.6 Gyr.  We, thus, need to further restrict the range of
models that are used.  In one extreme scenario, we could specify that
the SFR $e$-folding decay time is constant, say 0.6 Gyr, for all
galaxies. In this case, all the fitted models lie along one of the
curves in Figure~\ref{figFF}.  Consequently, the derived ages will
vary from 0.1 Gyr to many Gyr because the observed slopes $b$ take on
values spanning nearly the total range plotted in Figure~\ref{figFF}.
This was the basis for the modeling done in Paper II.
A somewhat more realistic scenario is to assume that all galaxies (in
a given cluster) have the same age and the star formation decay rate
is variable.  In this case, for clusters at $z$=0.9, the fitted models
fall along a horizontal line of the appropriate age in
Figure~\ref{figFF}.  The derived time constants will range from 0.2
Gyr, or less, up to values of 20 Gyr corresponding essentially to a
constant star formation rate with time.  One should note that there is
a region where slope $b$ is small and the ages are greater than 0.1
Gyr where no tau model will fit. Objects observed in this region must
have experienced a recent, large burst of star formation the
consequences of which are dominating the observed fluxes.
 
The fits of the model broadband photometry to the observations are
done using the maximum likelihood technique described in Paper II.
For constant age models, the fitting procedure yields a decay rate and
a goodness of fit indicator. Furthermore, the age of all galaxies is
assumed to be the time elapsed since the universe was 1.05 Gyr old
(for $H_o = 65$, $\Omega_o = 0.2$, $\Lambda_o = 0$ this corresponds to
$z = 6.0$). The best fit $e$-folding times (in Gyr) are listed in
column 9 in Tables 2, 3, and 4. If an acceptable fit is not possible
for $\tau < 20$ Gyr, the timescale is simply listed as ``long".  For
consistency with Paper II, where we compared our data only to constant
tau models, we also provide (in column 11) the age derived from
fitting our broadband BVRI data to the $\tau = 0.6$ family of models
(referred to as the ``color age'' in Paper II).

\subsection{Spectral Classification}
\label{sec:spclass}
 
The features in a galaxy's spectrum are age-dependent.
In particular, the metal lines are weak
in young hot objects and rapidly strengthen as the stellar population
becomes old.  The Balmer lines are fairly weak for young hot objects,
rapidly increase to a maximum when A-type stars dominate at ages of
about 1 Gyr, and then rapidly become weak thereafter.
An important diagnostic of the age and metallicity of a galaxy can thus
be gleaned from classification of its spectrum.
Our spectral classifications are based on the line strengths of various
metal and Balmer absorption features. We do not use equivalent widths
as a classification parameter because they tend to have large errors.
We do not use emission features in the classification and, thus,
differ from the approach used by Dressler et al.\ (1999) who use
the widths of the [OII] and H$\delta$ lines as their primary discriminants. 
We opt to use a {\it visual} spectral classification scheme 
because the experienced eye can assess the strengths of all the key features
simultaneously and, at the same time, judge the level of noise (including
complications such as poor night sky line subtraction).  
Our spectral classes are defined as follows: 
\begin{itemize}
\item type {\it ``a''} objects have spectra which are dominated by
Balmer lines, lack any g-band absorption, and have a CaII K
line strength which is no more than 25\% of the strength of the 
H + H$\epsilon$ line; 

\item type {\it ``a+k''} objects have strong Balmer lines, a
CaII K line that is approximately 50\% of the strength of 
the H + H$\epsilon$ line, and a detectable g-band line;  

\item type {\it ``k+a''} objects have a CaII K line that 
is equal in strength to H~+~H$\epsilon$, and have a 
g-band feature that is as strong as H$\gamma$, if H$\gamma$ is not filled with
emission;  

\item type {\it ``k''} objects have a CaII K line is stronger than
H~+~H$\epsilon$, have very strong $\lambda$3835 and g-band features,
and exhibit little or no H$\gamma$ absorption.
\end{itemize}
\noindent Our classes are given in column 10 in Tables 2, 3, and 4. 
Because the classification scheme above differs from the one used
by Dressler et al.\ (1999), it is important to give a rough translation
between the two methods. We achieve this by
comparing their sample spectra (Figure 5
of Dressler et al.\ 1999) with our templates, we find the following
correspondence:  Dressler {\it ``k''} is equivalent to our {\it
``k''};  Dressler {\it ``k+a''} is between our {\it "k"} and {\it
``k+a''};  Dressler {\it ``a+k''} corresponds to our {\it ``k+a''};
our {\it ``a+k''} is between Dressler {\it a+k"} and {\it ``e(a)''};
our {\it ``a''} corresponds to the remaining Dressler {\it ``e(a)"}.
Note that no classification is provided for galaxies 
where H$_{\gamma}$, H$_{\delta}$, and H$_{\epsilon}$ are in emission. 

\begin{figure}
\plotone{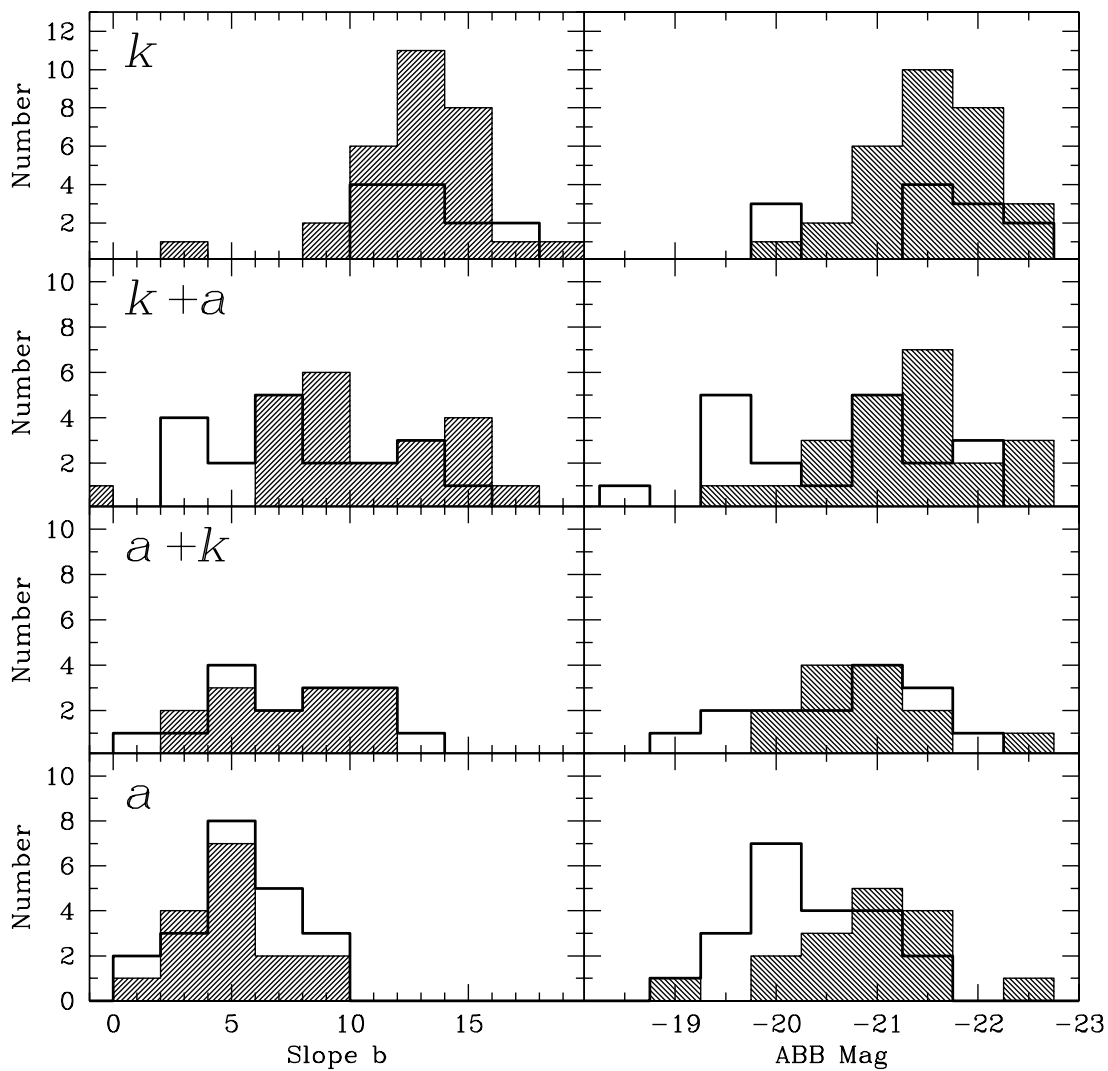}
\caption{The slope $b$ and ABB magnitude histograms as functions
of the spectral classification {\it a, a+k, k+a,} and {\it k} are shown.
The distributions for spectroscopically confirmed cluster members are
shown as shaded histograms. The histograms for field galaxies in
the range $0.65 \le z \le 1.1$ are unshaded.}
\label{figsptypevsb}
\end{figure}

Table 8 provides the specific mean values of equivalent widths and
luminosities for each spectral class.  Also given in this table, in 
parentheses, are the 1-sigma statistical errors in the mean values. 
In Figure~\ref{figsptypevsb},
we plot histograms of the slope $b$ and ABB magnitude for the four
spectral classes. The distributions for the confirmed members in all
three clusters are shown as shaded histograms. The unshaded histograms
show the distributions for field galaxies in the range $0.65 \le z \le
1.1$.  The mean slope $b$ value shows the expected qualitative
correlation with the spectral type increasing from $\bar b = 5.0$ for
class {\it a} to $\bar b = 12.9$ for class {\it k}. In other words,
the older, more metal rich galaxies tend to be, on average, redder.
The {\it k} type galaxies also tend to be significantly brighter (by
0.7 - 1 mag) than {\it a} type galaxies, a trend most likely
attributed to the dominance of massive elliptical galaxies in the {\it
k} spectral class. These results are consistent with those reported
by Poggianti \etal (1999).

\subsection{Emission Line Equivalent Widths and Intensities}
\label{sec:emeqw}

Emission line equivalent widths are measured as described in Paper
II.  As our sample is dominated by galaxies at high redshift, the
spectral region containing the [OII] line is nearly always observed.
The ${\rm H}\beta$ and [OIII] regions are also observed for the
majority of objects with $z \simless 0.95$, but at values of $z$ above
0.8 the signal is low and the night-sky emission bands are very
strong.  In a few low redshift galaxies the $H\alpha$ region is
observed.  Equivalent widths in Angstroms, corrected to the
rest-frame, have been measured for [OII] $\lambda3727$, H$\beta$, and
[OIII] $\lambda5007$.  These emission lines are often strong and are
well isolated from other spectral features. Emission line equivalent
widths are given as negative values (positive values are used for
absorption features).  The equivalent width of ${\rm H}\beta$ can be
either negative or positive since the line lies on top of the possible
absorption feature.  The typical errors in the equivalent widths are
estimated from the differences between rest equivalent widths derived
from two independent and comparable spectra of the same object.  For
the current sample, we have a total of 54 such overlap observations
with $0.5 \simless z \simless 1.0$.  The comparisons suggest errors of
$\sim5$\AA\ for the [OII] line and $6-8$\AA\ for the H$\beta$ and
[OIII]$\lambda5007$ lines.  The rest equivalent widths in Angstroms of
[OII], ${\rm H}\beta$, and [OIII] are given in columns 12 through 14
in Tables 2, 3, and 4.  No corrections for underlying absorption have been
applied to the ${\rm H}\beta$ equivalent widths listed in the tables.

Emission line intensities are calculated by measuring the observed
flux of the continuum at the line and converting the equivalent widths
to intensities. For the ${\rm H}\beta$ line, this conversion is

\begin{equation}
{\rm log} (I({\rm H}\beta)) = {\rm log} (EW({\rm H}\beta))-0.4
AB(4861)_{absol} + 31.74
\end{equation}

\noindent where $AB(4861)_{absol}$ is the observed AB magnitude of the
continuum measured at the redshifted position of H$\beta$, converted
to an absolute magnitude using the distance modulus. The intensity
$I({\rm H}\beta)$ is in units of erg s$^{-1}$. Similar equations apply
to the [OII] and [OIII] lines: the zeropoints are 31.97 for the [OII]
line and 31.72 for the [OIII] line. The continuum is interpolated
using the broadband AB magnitudes and the fitted models.
In these intensity calculations a correction for the equivalent width
of the absorption line underlying H$\beta$ {\it has} been applied
based on the spectral synthesis model being compared to the data.  The
correction is typically 3 to 8\AA.

We validate our line intensity measurements by comparing line ratios
to previously published work. Specifically, we use the best spectra in
the appropriate redshift range to derive a mean observed intensity
ratio of [OII] to H$\beta$ and [OIII] to H$\beta$.  Averaging the
logarithms of the observed intensity ratios we find $I([{\rm OII}])/
I({\rm H}\beta) = 2.19$ and $I([{\rm OIII}])/ I({\rm H}\beta) = 1.81$.
Because the redshifts are high, and the galaxies are small on the sky,
these represent averages over the whole galaxy.  Kennicutt (1992) has
measured these ratios in nearby, mostly late-type galaxies using large
apertures and small telescopes. His better data give ratios of 2.40
and 1.48, respectively. These lines have been measured by van Zee
\etal (1998) in many HII regions over the whole surfaces of a variety
of spiral galaxies.  After removing their corrections for reddening,
their average ratios are 2.14 and 1.09, respectively.  The intrinsic
scatter is about 50\% for $I([{\rm OII}])/ I({\rm H}\beta)$ and about
a factor 2 for $I([{\rm OIII}])/ I({\rm H}\beta)$.  Our results agree
quite well, given the scatter, with the results of Kennicutt and Van
Zee \etal although our average [OIII] to H$\beta$ ratio is higher than
that in those two studies.

\subsection{Metal-Line Equivalent Widths}
\label{sec:mleqw}

Although we have measured eight absorption lines in the galaxy
spectra, they are not all equally useful.  Among the metal lines, the
$\lambda3835$ feature is in a spectral region where any continuum
choice is rather arbitrary.  The feature is strong in the redder
objects but it is superposed on the Balmer H9 line, which will be
relevant for bluer galaxies. The CaII H line is superposed on
H$\epsilon$. The g-band is close to H$\gamma$ but sufficiently
separated from it to be measurable; it is usually sufficiently far in
the red to be in a rather noisy spectral range.  Of the available
Balmer lines, H$\gamma$ is always suspect since it may be partially or
completely filled with emission.  H$\delta$ should be reliable. H8 is
in a crowded spectral region where a continuum is very hard to
estimate.

\begin{figure}
\plotone{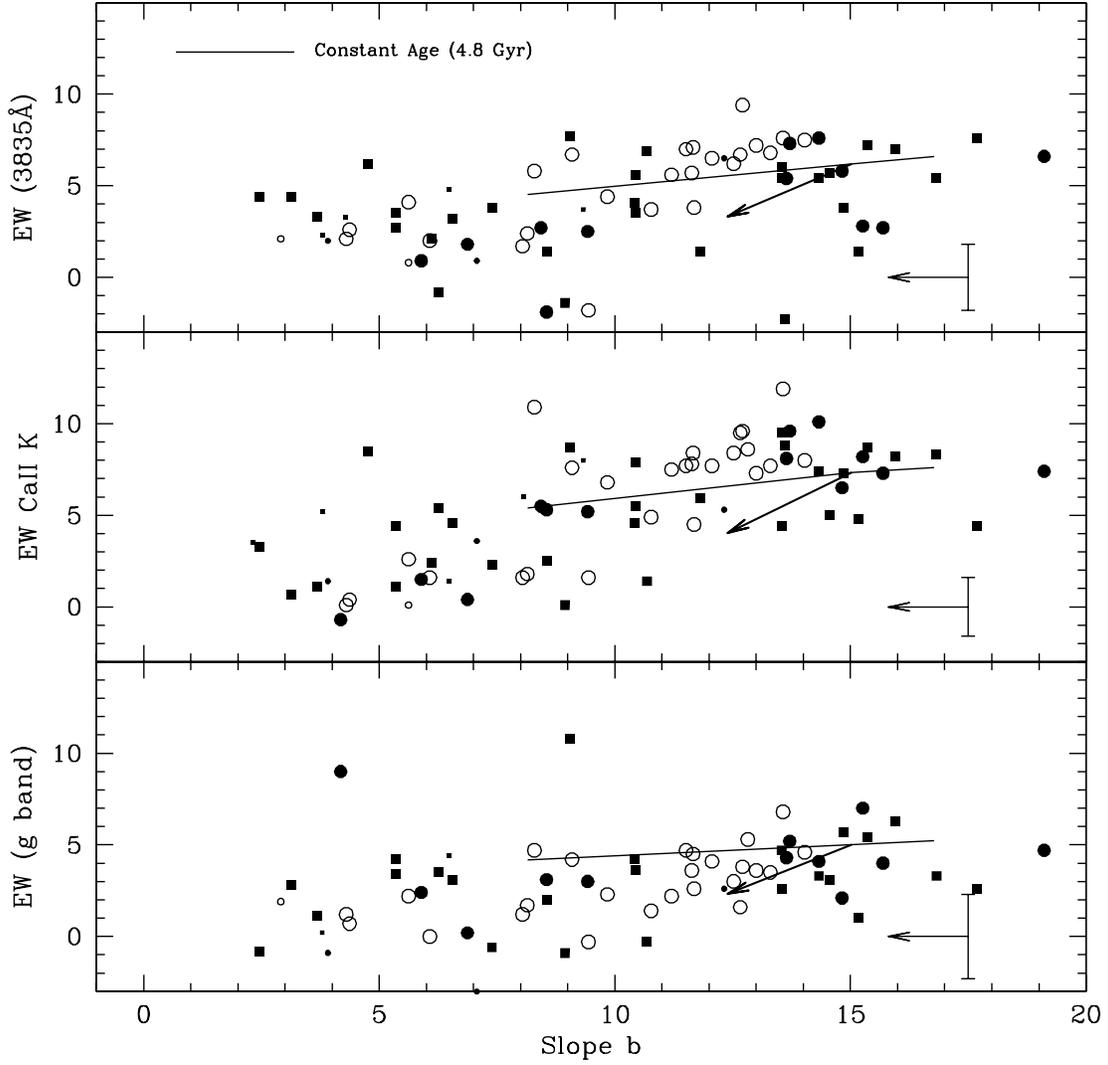}
\caption{The relation between the rest equivalent width (in Angstroms)
of the $\lambda$3835, CaII K, and g-band spectral features and the
slope $b$ for confirmed cluster members.  The predictions from the
Bruzual \& Charlot (1993) solar metallicity model with a constant age
(4.8 Gyr at $z = 0.90$) are shown by the solid line.
The open circles, filled circles, and filled
squares display the observations (uncorrected for reddening) for
\clalpha, \cldelta, and \cleps, respectively. The larger symbols
represent results based on very high S/N spectra.
A typical observation error bar is shown.
The observations are shown uncorrected for reddening.
The arrow on the error bar shows how
the data would shift if corrected for a reddening of $E(B-V) = 0.2$
mag. The diagonal arrow on the model curve shows how the model shifts
when the metallicity is reduced to 0.2 solar.}
\label{figD}
\end{figure}

The rest equivalent widths of the $\lambda$3835 feature, CaII K and
the g-band are plotted against the continuum slope $b$ in
Figure~\ref{figD} for cluster members. 
In this figure, and in subsequent similar ones, cluster members in the
three clusters \clalpha, \cldelta, and \cleps\ are indicated by open
circles, filled circles, and filled squares, respectively.
The errors in the highest S/N
equivalent width measurements of the $\lambda3835$ line, the CaII K
line, H$\delta$, and the g-band are, respectively, 1.8, 1.6, 3.0, and
2.3\AA. The mean equivalent widths as a function of slope $b$ for the
sub-sample of red and blue galaxies with the best spectra are given in
Table 8.

To compare the absorption-line data with the models it is desirable to
use models with approximately the same spectral resolution as the
observations. The BC96 model spectra have a resolution or binning size
of 20\AA\ which is twice as large as the spectral resolution of our
observations making direct comparisons with the data difficult.  The
1993 Bruzual \& Charlot (BC93) models, on the other hand, have a
resolution of 10\AA\ and are, thus, ideal for comparisons with
observational data.  Furthermore, a comparison of the Bruzual \&
Charlot 1993 and 1996 models with the bandpasses optimized for the
1996 models shows that the absorption equivalent widths are larger for
the 1996 models by factors that vary from 1.27 to 1.69 depending on
the line being measured. That the factor varies from line to line
strongly suggests that the mismatched spectral resolution is indeed at
least part of the source of the discrepancy.  Further evidence in
support of this explanation comes from the excellent agreement between
the equivalent widths obtained from a very high S/N spectrum of NGC
4889 in the Coma cluster\footnote{The NGC 4889 spectrum was obtained
with the Double Spectrograph on the Hale 5m telescope at Palomar
Observatory.}  and the BC93 solar metallicity models.  We therefore
make comparisons of observed and measured line equivalent widths using
the BC93 models.

The line strengths from the BC93 solar metallicity tau0.3, tau0.6, and
tau1.0 models, all with an age of 4.8 Gyr which is close to that
expected for cluster galaxies at $z \sim 0.9$, are shown in
Figure~\ref{figD} as the solid curves.  The figure demonstrates that the
BC93 equivalent widths are in very good agreement with the
corresponding values for our distant cluster galaxies.

\subsection{The Balmer Jump and the 4000\AA\ Break}
\label{sec:jump}

\begin{figure}
\plotone{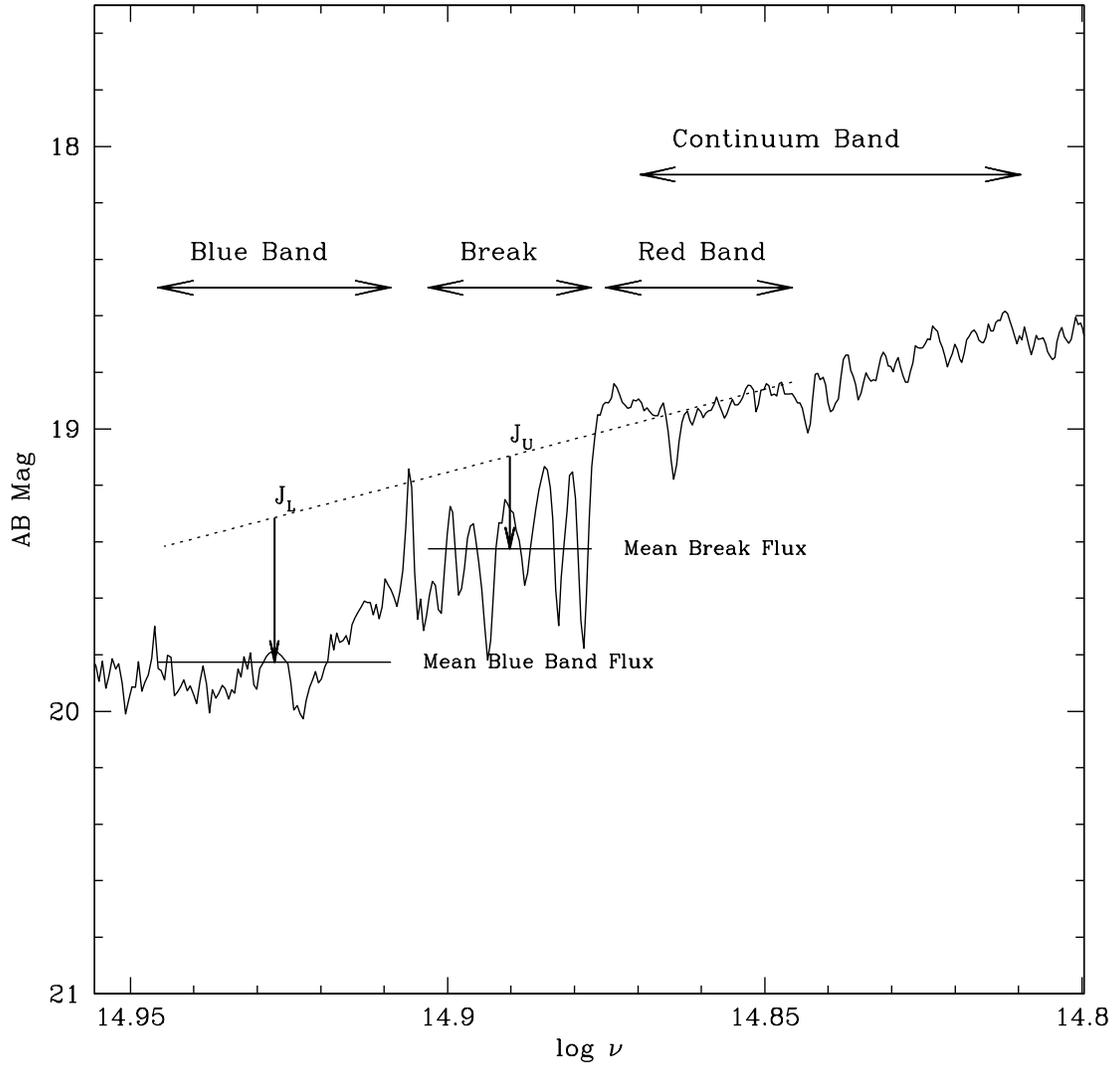}
\caption{A visualization of the definitions of the $J_l$ and $J_u$
jump amplitudes. The dashed line is the extrapolation of the continuum
level, normalized to the mean red band flux. The jump amplitudes are
measured with respect to this extrapolation.}
\label{figjmpdemo}
\end{figure}

The conventional 4000\AA\ discontinuity measure, D, is not
particularly useful for the analysis of young stellar systems (like
those in distant, late-type galaxies) because it does not enable one
to discriminate between spectra that are dominated by metal-line
absorption from cool stars and by spectra dominated by the high Balmer
lines and the Balmer jump in A-type stars.  Furthermore, the
measurement D does not remove the overall color of the object that
depends on both the spectral energy distribution and on the wavelength
response of the detector (because our spectra are not flux
calibrated).  We prefer instead to use the $J$ parameter (first
proposed in Paper II).  For the present analysis, we define two
slightly modified jump parameters, $J_l$ and $J_u$.  Continuum points
are computed by first establishing the mean slope of the spectrum in
the rest-frame spectral range 4050-4650\AA, which is usually much
better exposed than the region below 3700\AA. Points near H$\delta$
(4075\AA\ to 4125\AA) and the g-band and H$\gamma$ (4280\AA\ to
4350\AA) are excluded from the calculations.  The level of this
continuum is normalized to match the mean AB flux in the range 4000 to
4280\AA. This normalized continuum level and the measured slope are
then used to extrapolate a continuum in the 3750 to 3980\AA\ and 3400
to 3700\AA\ regions.  Figure~\ref{figjmpdemo} provides a schematic
description of the jump definitions.

The jump $J_l$ is defined as the difference of the average flux
(expressed in AB magnitudes) measured from 3400 to 3700\AA\ and the
defined continuum at the central wavelength of this band (\ie
3550\AA).  The jump $J_u$ is defined as the difference of the average
flux (again in AB magnitudes) in the range 3750 to 3980\AA\ and the
defined continuum at the central wavelength of 3865\AA.  $J_u$ is very
similar to the conventional jump D except that the continuum slope has
been removed. It essentially measures the strength of the
discontinuity at 4000\AA\ caused by metal and late Balmer series
absorption lines.  $J_l$ measures the Balmer jump produced by hydrogen
absorption plus any metal-line blanketing in the 3400 to 3700\AA\
region.  Since both $J_l$ and $J_u$ depend on an extrapolation of the
continuum that is defined above 4000\AA, they are subject to
systematic errors caused by loss of light due to refraction and
absolute calibration errors in the blue relative to the red spectral
region.  This effect should be small for $J_u$ but can be large for
$J_l$.  Subject to these systematic errors, the ratio of $J_u$ to
$J_l$ is a measure of the relative importance of the Balmer jump and
the 4000\AA\ break.  If the 4000\AA\ break is large, both $J_u$ and
$J_l$ will be close to unity.  If the Balmer Jump dominates, $J_u$
will be less than half of $J_l$.  The measured values of $J_u$ and
$J_l$ are given in columns 15 and 16 in Tables 2, 3, and 4. Errors
have been estimated in the same manner as those for the equivalent
widths.  The typical errors in $J_u$ and $J_l$ are 0.10 mag and 0.13
mag, respectively.

\begin{figure}
\plotone{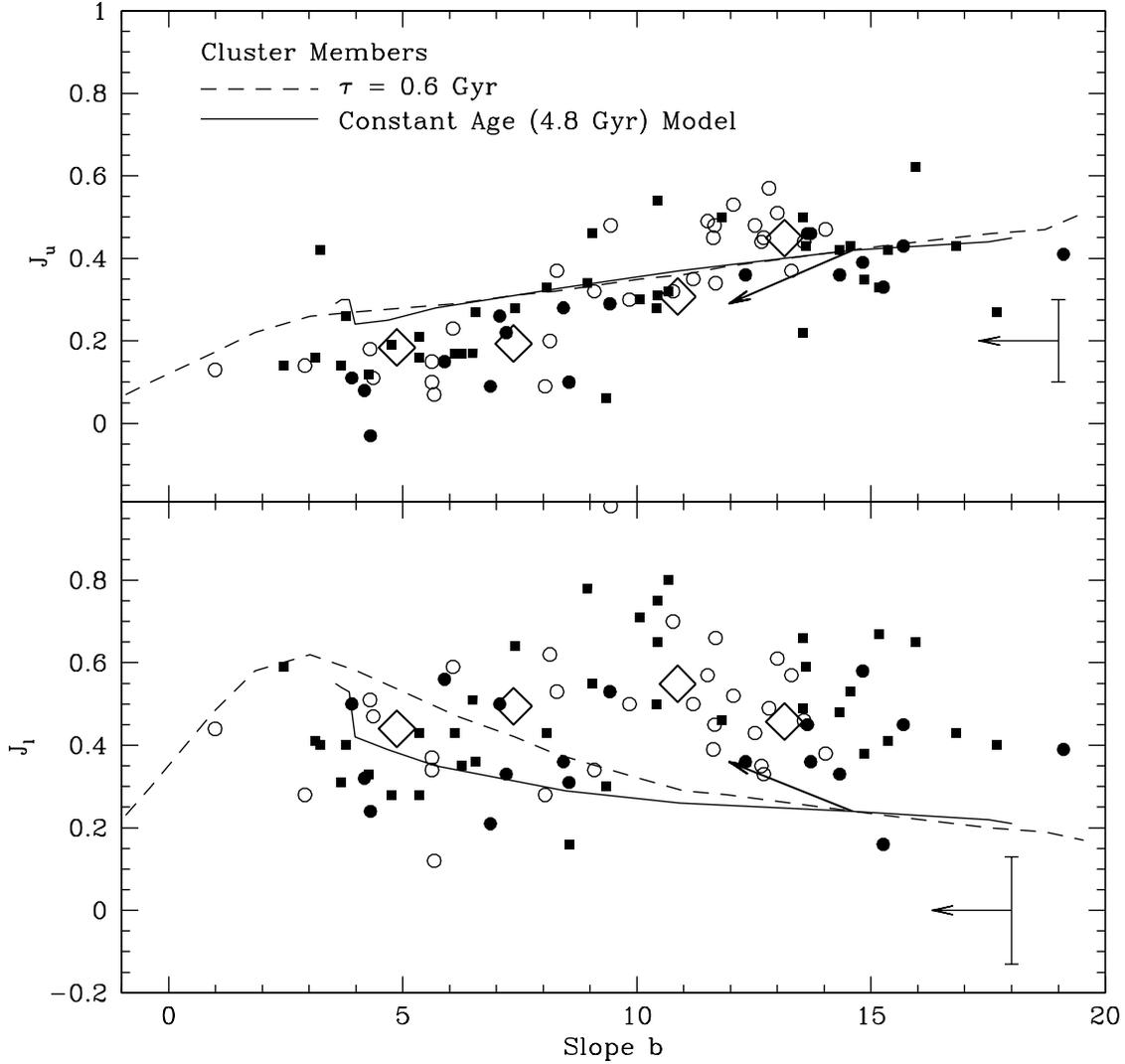}
\caption{The upper and lower jumps ($J_u$ and $J_l$) in AB magnitudes
as functions of the slope $b$ for galaxies in the clusters are shown
in the upper and lower plots, respectively.  The symbols and arrows
are as defined in Figure~\ref{figD}.  Typical error bars are shown.
Large open diamonds show mean values in four bins for the highest
quality spectra.  The broken curves are for tau0.6 solar metallicity
BC96 models of various ages with redshift $z = 0.9$.  The solid curves
are for constant age models.
A typical observation error bar is shown.
The observations are shown uncorrected for reddening.
The arrow on the error bar shows how
the data would shift if corrected for a reddening of $E(B-V) = 0.2$
mag. The diagonal arrow on the model curve shows how the model shifts
when the metallicity is reduced to 0.2 solar.}
\label{figjujl}
\end{figure}

The observed values of $J_u$ and $J_l$ are plotted against the
observed slope $b$ for members of the three clusters in
Figure~\ref{figjujl}.  Representative error bars are shown.  The large
open diamonds show the mean jump values in four slope $b$ bins for the
highest quality spectra only. These mean values are consistent with
the full data distribution, indicating that no serious systematic
errors are introduced into the jump estimates at lower signal-to-noise
levels.

The solid curve shows the jump values predicted by our constant age
scenario models.  The dashed curve shows the predictions from a solar
metallicity tau0.6 model.  The small discontinuity in the constant age
model predictions at small slope $b$ values appears to be caused by
some difference in the tau5.0, tau10.0, and tau20.0 models or in our
interpretation of the models.

The solar metallicity models do a reasonable job of reproducing the
observed correlation between $J_u$ and the slope $b$.  There is no
significant observed correlation between $J_l$ and the slope $b$.
This is, in part, due to the fact that $J_l$ is affected both by the
Balmer jump and the 4000\AA\ break which tend to evolve in a manner
which minimizes the change in $J_l$. Furthermore, the absolute values
of $J_l$ are subject to large systematic errors (more than $J_u$), and
the difference between the observations and models is not significant
when these errors are taken into account.

\subsection{Metallicity and Reddening}
\label{sec:zred}

To obtain rough constraints on the metallicities of the galaxies in
our survey, we must rely on comparisons with the BC96 models as the
BC93 models do not include non-solar models. In the following, we
therefore use the measured BC96 equivalent widths scaled to correct
for the spectral resolution difference between the BC96 models and
LRIS data.  The vectors drawn in Figures~\ref{figD} and~\ref{figjujl}
show how the reddest models shift when the metallicity is changed from
solar to 0.2 solar. The shift that is shown is typical when the slope
$b$ is greater than 6. At small $b$, the shift is smaller because the
metal lines have diminished in strength.  The observed CaII K and
the$\lambda$3835 features show excellent consistency with the solar
metallicity models.  The g-band measurements would be fitted better
with slightly sub-solar metallicity models. The uncertainties,
however, are too large to be definitive.  The J$_u$ data suggest that
a metallicity slightly less than solar would fit the data best. For
J$_l$ the fits of the models to the observations are all poor. This
may indicate that a substantial amount of the blue light from the
galaxy is being lost at the slit.

Corrections for reddening will change the slope $b$ but not the jumps
or equivalent widths.  The horizontal arrows on the representative
error bars in the above figures show how the data points would shift
if a 0.2 mag de-reddening correction were applied. A reddening of 0.2
mag in $B-V$ corresponds to a change in slope $b$ of 1.7. Such a
reddening correction would not significantly alter the fits to the
models. A much larger reddening correction, however, would make the
fits of the equivalent widths unacceptable.

\section{Cluster Properties}
\label{sec:clusprop}

\subsection{Brightest Cluster Galaxies}
\label{sec:bcg}

\begin{figure}
\plotone{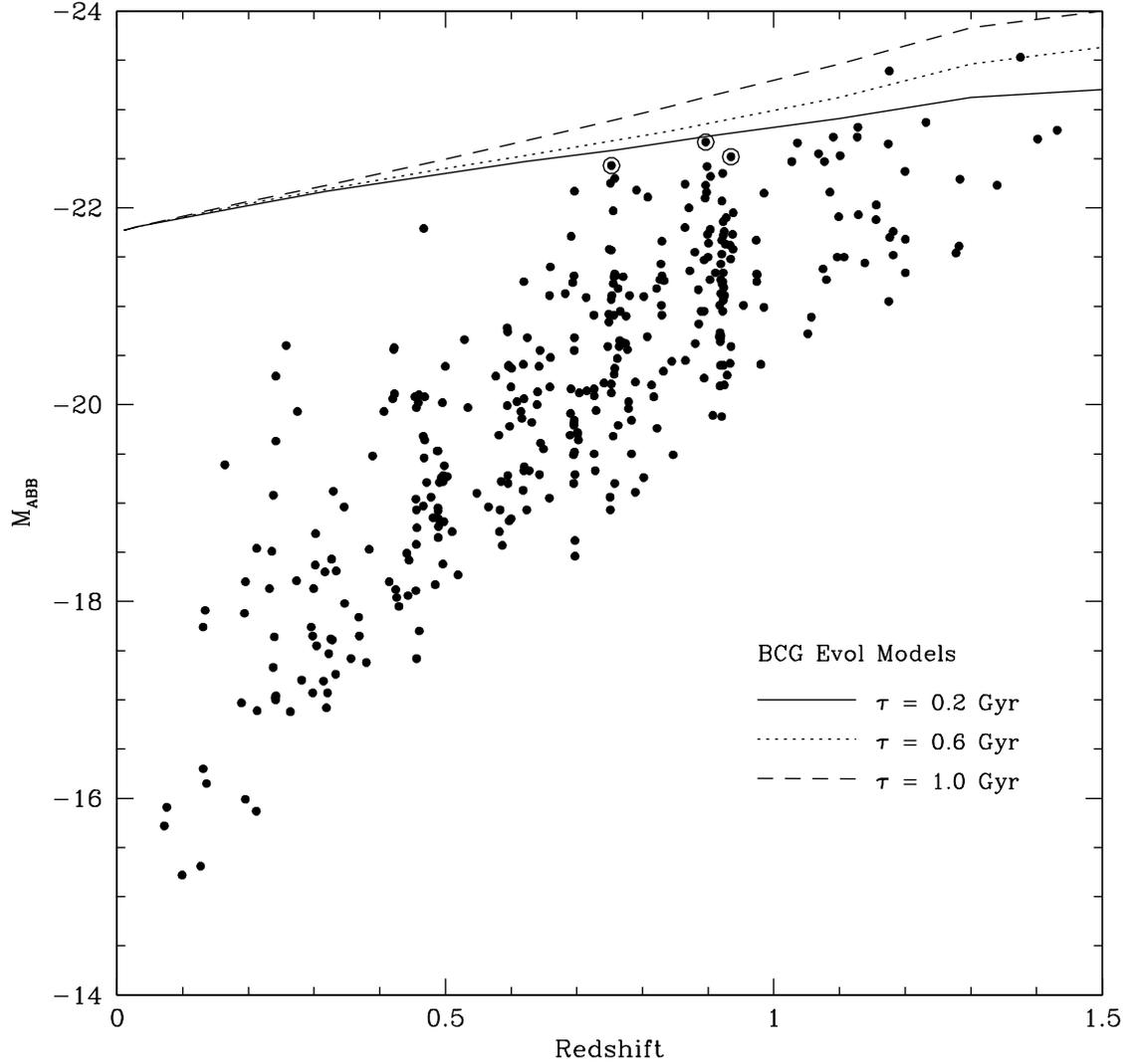}
\caption{The absolute AB magnitude in the rest frame $B$ band as a
function of redshift for all galaxies with measured redshifts. The
data for the brightest cluster galaxy for each cluster is highlighted
with a large open circle. The redshift evolution of the BCG magnitude
for the tau0.2, tau0.6, and tau1.0 models are also shown.}
\label{figR}
\end{figure}

The absolute AB magnitudes of the galaxies in the rest B band,
$M_{ABB}$, have been calculated as described in \S2.2.2 of Paper II
and are listed in column 18 of Tables 2, 3, and 4.  Figure~\ref{figR}
shows $M_{ABB}$ plotted against the redshift $z$. The lower boundary
of points is a consequence of the survey magnitude limit.  The upper
boundary at lower values of $z$ is strongly influenced by the small
volume sampled at low redshifts.

The $M_{ABB}$ magnitudes predicted by the tau0.2, tau0.6, and tau1.0
models are listed in Table 9 as a function of $z$ and are plotted in
Figure~\ref{figR}. (The results for ssp and tau0.2 models are almost
identical since there is virtually no star formation and evolution is
proceeding slowly.)  The constant age scenario is assumed, that is,
star formation commenced 1.05 Gyr after the universe began.  The
curves are normalized to match the $z \le 0.05$ brightest cluster
galaxy (BCG) sample of Postman \& Lauer (1995) which, after correcting
their results to conform to $H_o=65$ and a metric radius of 22.3\ih65
kpc (to match our aperture photometry at $0.76 \simless z \simless
0.92$), and adopting a mean $\alpha$ parameter value of 0.50, gives
$M_{ABB,BCG}= -21.81$ at $z \le 0.05$.

Changing the star-formation-rate decay time from 0.2 to 1.0 Gyr yields a
significant change in the BCG luminosity, $\sim0.5$ mag by $z = 1$,
since star formation becomes increasingly important. The median value of 
tau for our constant age models is 1.0 Gyr.   
For the same models,  the result of
changing from $H_o=65$ to $H_o=80$ is shown in Table 9.  This change makes 
a relatively small difference, even at high $z$ ($\simless 0.2$ mag by 
$z = 1.5$). The BCG in \clalpha, Keck \#2151, has 
$M_{ABB} = -22.44$.  The best fit constant age 
scenario model gives a decay time of $\tau = 0.9$ Gyr and predicts 
$M_{ABB} = -22.84$. The BCG in \cldelta, Keck \#1888, has 
$M_{ABB} = -22.51$. The best fit model gives $\tau = 0.6$ Gyr and predicts 
$M_{ABB} = -22.86$. The BCG in \cleps, Keck \#1292, has 
$M_{ABB} = -22.52$. The best fit model gives $\tau = 0.9$ Gyr and predicts 
$M_{ABB} = -23.06$. On average, the models predict a BCG magnitude that is 
$\sim 0.4$ mag brighter than observed.  

\subsection{The ``Red Envelope"}
\label{sec:redenv}

\begin{figure}
\plotone{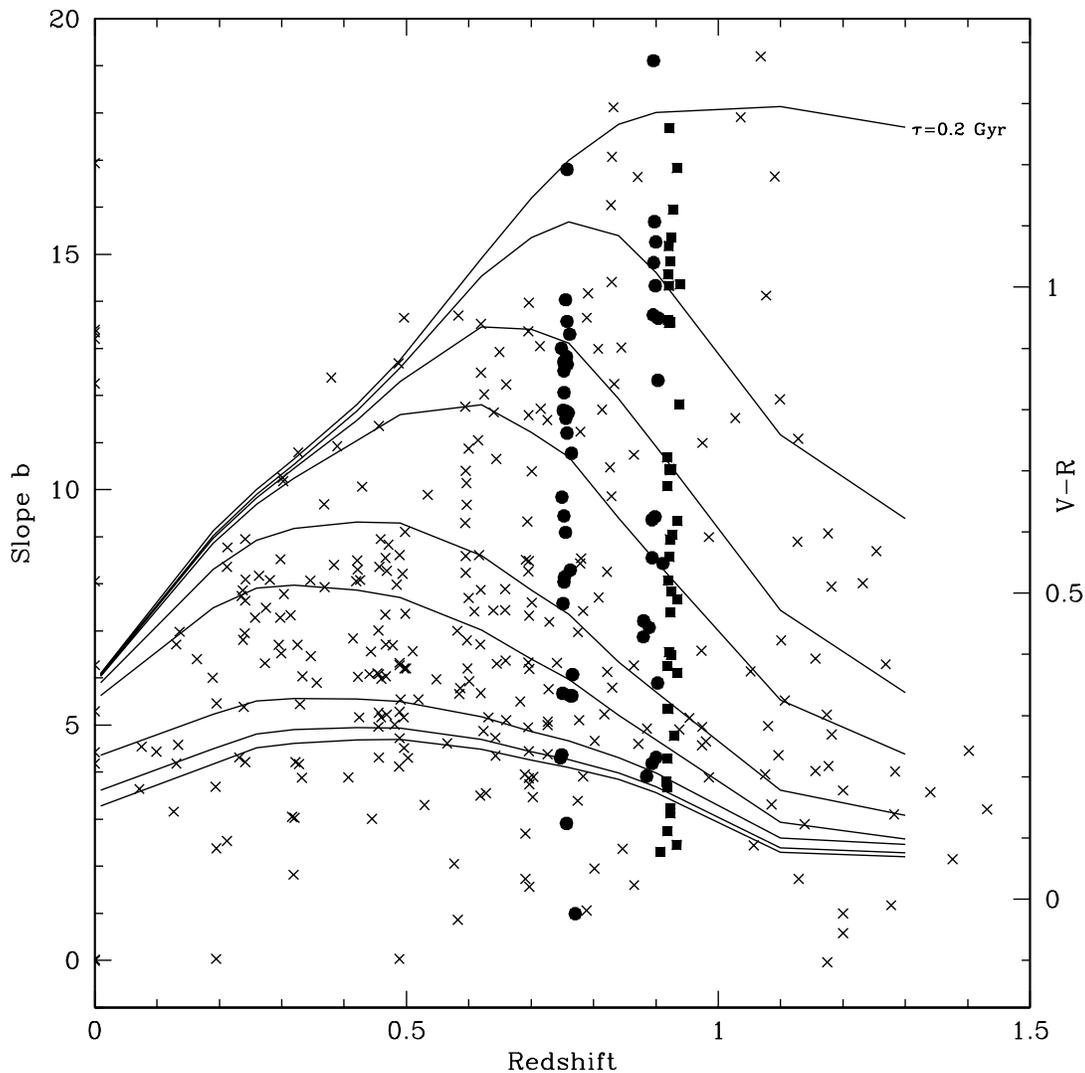}
\caption{A plot of the slope $b$ versus redshift. Cluster members are
shown as solid points, field galaxies as crosses.  The relationship
between age and redshift are calculated based on a Friedman model with
$H_o=65$ and a value of $q_o=0.1$. It is assumed that star formation
commenced 1.05 Gyr after the universe began.  Each curve represents
the predicted evolution of a galaxy that follows various tau models. From
top to bottom the curves correspond to tau0.2, tau0.6, tau0.8,
tau1.0, tau1.5, tau2.0, tau5.0, tau10.0, and tau20.0. The righthand
ordinate is an approximate $V-R$ color scale.}
\label{figS}
\end{figure}

One of the striking features of the galaxies in our three samples is
the large spread in the color of the cluster members, as measured by
the slope $b$.  This is dramatically demonstrated in
Figure~\ref{figS}.  As in Figure~\ref{figCC}, there is almost no sign
of the red color ridge that is often seen in clusters at $z = 0.5$ (\eg
Oke, Gunn, \& Hoessel 1996; Stanford, Eisenhardt, \& Dickinson 1998,
hereafter SED98) although there are still quite a number of red
galaxies in the samples. The visibility of the red ridge at $z = 0.9$
is a bit more prominent when a near-infrared color is used, as the CM
diagrams of SED98 demonstrate.  All of the objects in the $z = 0.9$
clusters with slope $b$ greater than 13 to 14 correspond to the
central concentration of red galaxies. In \clalpha\ ($z = 0.76$) the
galaxies populating the red ridge have $b$ greater than 11. The broad
color distribution of cluster galaxies at $z \simgreat 0.8$ is
compatible with the evolution predicted from our family of tau
models. This is demonstrated by the curves in Figure~\ref{figS} which
show how the $b$ slope of galaxies evolves with redshift.  The nine
curves from bottom to top correspond to the BC96 tau20.0, tau10.0,
tau5.0, tau2.0, tau1.5, tau1.0, tau0.8, tau0.6, and tau0.2 families of
models.  Specifically, the model spectra are allowed to age by times
that are calculated assuming $H_o=65$, $\Omega_o=0.2$, and by assuming
that star formation commenced 1.05 Gyr after the universe began.  As
can be seen, all of the objects at $z$=0.9 with values of $b$ between
7 and 19 have collapsed into a much narrower range from 11 to 12.5 by
$z$=0.5, corresponding to a range in $V-R$ or $R-I$ of 0.1 mag.  The
red ridge is, thus, predicted to contain a much larger fraction of the
galaxies in a cluster at $z = 0.5$ than at $z = 0.9$, making the envelope
much more prominent at the lower redshifts.  This is clearly what is
seen in our the results and those from SED98.  The bluest objects at
$z$=0.9 remain very blue by $z$=0.5, however.  The set of models shown
in Figure~\ref{figS} encompass nearly all the galaxies seen at all
redshifts.  There are a few objects that are even bluer than our most
extreme model.  These are presumably objects where there has been a
recent very strong burst of star formation.

\section{Emission Line Intensities and Star Formation Rates}
\label{sec:emlsfr}

\subsection{Fraction of Active Galaxies}
\label{sec:active}

In Paper II we demonstrated the strong evolution in the fraction of
active galaxies, defined to be those galaxies with [OII] rest
equivalent widths greater than 15\AA, in the field and in clusters at
$z > 0.7$ relative to what is seen at the present epoch. We now
provide an improved constraint by including the results from \clalpha\
and \cleps.  In total there are 345 galaxies with redshifts and 25
additional galaxies without redshifts but that have sufficiently high
S/N spectra that we can be fairly sure there are no emission lines.
Ten objects with low redshifts have strong ${\rm H}\beta$ and [OIII]
but [OII] lies bluewards of the observed spectral range. For field
galaxies with $0.40 < z \le 0.85$, 65\% are active. For field galaxies
with $z > 0.85$, the fraction is 79\%.  For roughly the same range in
redshift, Hammer \etal (1997) find quite comparable values of 65\% and
90\%.  Within the central 1.5 \ih65 Mpc regions of our three distant
clusters, the fraction of active galaxies is 45\%. This is
substantially higher than the 10 to 20\% active galaxy component seen
in the centers of $0.2 < z < 0.55$ clusters (Balogh \etal 1997) but
significantly lower than the active fraction in the surrounding field.

\begin{figure}
\plotone{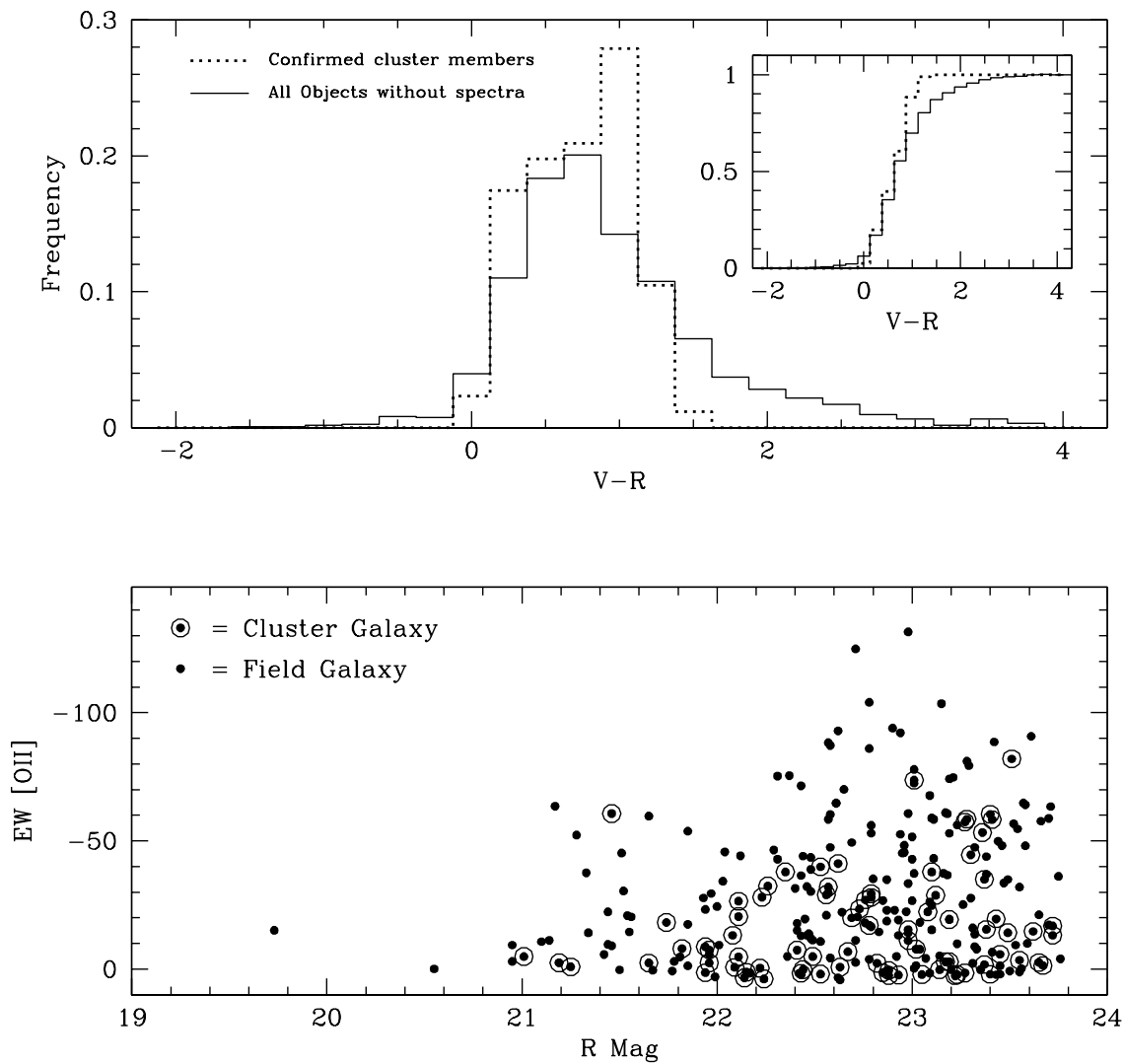}
\caption{Lower plot shows the measured [OII] equivalent
width as a function of galaxy $R-$band magnitude. Cluster members are
shown as circled points. Galaxies defined as active ([OII] EW greater than 15\AA)
are distributed across a broad magnitude range. Upper plot shows the
$V-R$ color distributions for the spectroscopically confirmed cluster
members and for objects without spectroscopic data (within the same
apparent magnitude range as the cluster members).}
\label{figo2vsmag}
\end{figure}

We must examine whether or not our $R-$band selection criterion biases
our estimate of the active galaxy fraction. The [OII] line redshifts
into the $R$ passband at $z \approx 0.53$ and redshifts out of the
$R$ passband at $z \approx 1.15$. Hence, one could imagine that selecting
targets for spectroscopy based solely on the object's $R-$band magnitude
might potentially bias the sample in favor of objects with strong
[OII] emission near the survey magnitude limit and, consequently, cause
the active fractions to be overestimated. Our selection criterion, however, does not
appear to have a significant effect. We demonstrate this in two ways. First, the lower
plot in Figure~\ref{figo2vsmag} shows the [OII] equivalent width 
as a function of the galaxy $R-$band apparent magnitude.
There is no preference for galaxies with equivalent widths
greater than 15\AA\ to lie near the survey magnitude limit. 
Second, the upper plot in Figure~\ref{figo2vsmag} shows the $V-R$ color distributions for
the 91 spectroscopically confirmed cluster members (in all three clusters)
and for all 1239 objects without spectroscopic observations that have
$R$ magnitudes in the {\it same} range as the confirmed cluster galaxies.
A Kolmogrov - Smirnov test test indicates that
these two distributions are only marginally inconsistent -
the hypothesis that the two distributions
are drawn from the same parent population is rejected at the 90\% confidence level. 
This is less than a 2$\sigma$ difference. The source of this small difference
is a red tail ($V-R \ge 1.75$) in the $V-R$ distribution in the objects without spectra.
As shown in Figure~\ref{figDD} (see below), there is a rough correlation
between [OII] equivalent width and galaxy color.
If the objects that were not targeted for spectroscopy were preferentially low [OII]
emitters, one might expect the distribution of $V-R$ colors in the unmeasured
sample to contain a higher fraction of red objects than the spectroscopic sample.
The objects with $V-R \ge 1.75$ comprise $\sim13$\% of all unobserved objects.
However, these reddest objects are distributed relatively uniformly across the
imaged areas and are not clustered about the centers of the clusters. It is thus
unlikely that they are all cluster members. Only 10\% of the spectroscopically
unmeasured objects that lie within 500\ih65 kpc of the cluster centers
are redder than $V-R = 1.75$. A conservative limit
to the amplitude of any systematic error in our active fraction would thus
be about a $\sim10$\% overestimate. The actual systematic error is likely
to be significantly smaller.
We conclude that our estimates of the active fractions in the field
and in the clusters are, thus, not strongly biased by our spectroscopic target
selection process.

Star formation in galaxies manifests itself in the rest optical
bandpass in two important ways.  First, star formation leads to the
formation of HII regions which, in turn, generate emission lines and,
in particular, the [OII], [OIII], and Balmer lines.  Second, star
formation generates a population of hot stars that make the observed
energy distributions of the galaxies blue. The slope $b$ parameter and
the emission line equivalent widths are quantitative measures of star
formation. Specifically, the slope $b$ measures the importance of
young stars in the spectrum.  The equivalent width is the ratio of the
star formation rate during the last few million years to the continuum
generated by the main sequence over the total history of the
galaxy. The rest equivalent widths of the [OII] line as a function of
the overall spectral energy distribution as represented by the slope
$b$ are displayed in Figure~\ref{figDD} for spectroscopically
confirmed members of the clusters \clalpha, \cldelta, and \cleps. 
Qualitatively, the results are as expected: galaxies with strong
emission are blue (small $b$) while objects with very little emission
are red (large $b$).  The predicted $b$ -- [OII] EW relationships for
the constant age (4.8 Gyr) and the tau0.6 models are shown as well for
$z = 0.90$.  For the constant tau models, the sensitivity of the
relation to the adopted decay rate of star formation is small, largely
because the presence of hot stars and [OII] emission are somewhat
coupled. In fact, the curves for tau = 0.2 to tau = 3.0 are almost
identical.  The models fit the majority of these observations quite
well suggesting that the constant age scenario (with variable star
formation decay rates) is, on average, a plausible hypothesis. The
constant tau models also provide a good fit.  Galaxies which lie to
the upper right of the curves are presumably those systems that have
undergone a recent burst of star formation.

\begin{figure}
\plotone{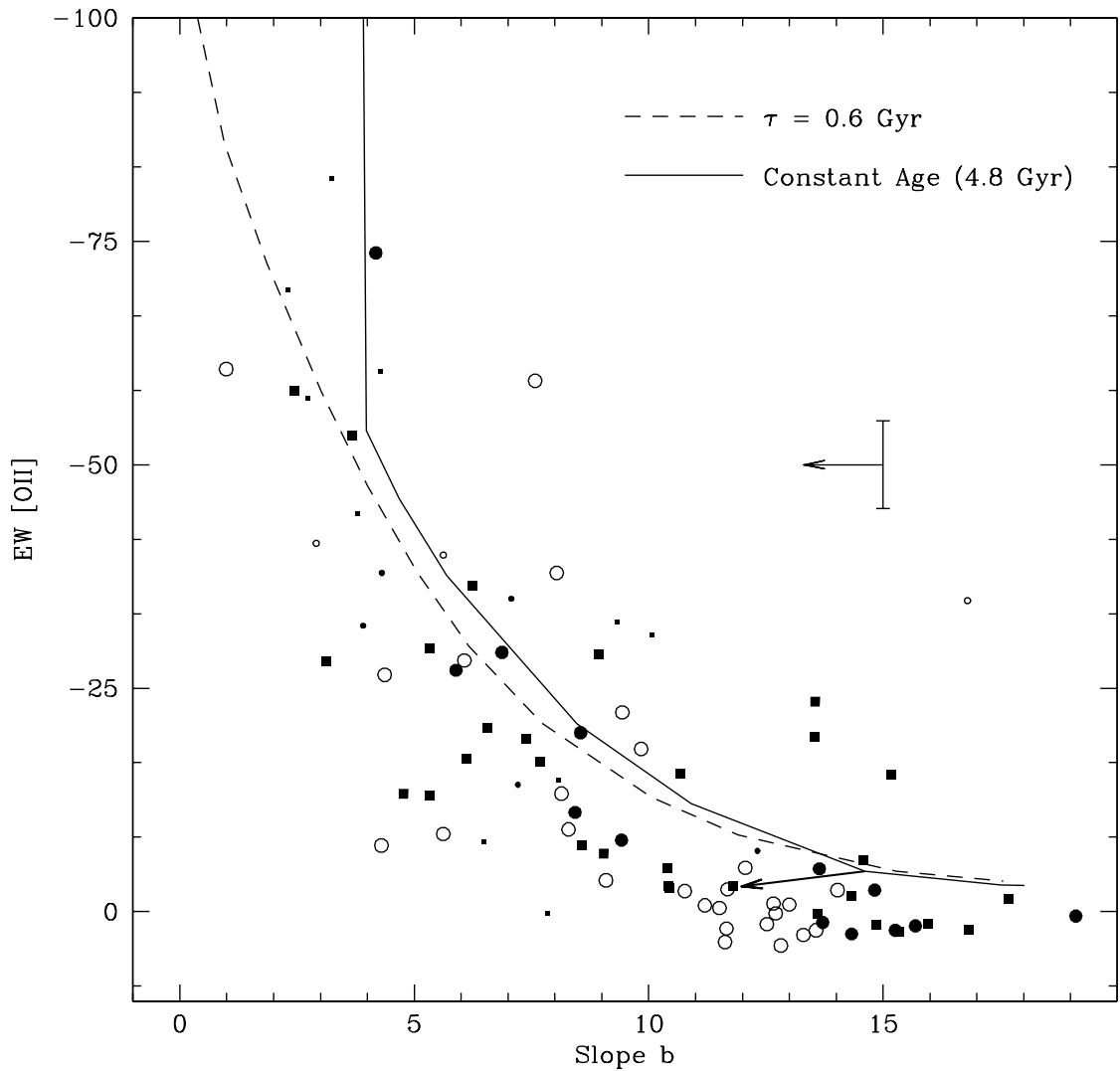}
\caption{The relation between the rest equivalent width (in Angstroms)
of the [OII] line and the slope $b$ for confirmed cluster members.
The predictions from the Bruzual \& Charlot 1996 spectral synthesis
package for a solar metallicity model with a constant age (4.8 Gyr at
$z = 0.9$) is shown. The symbols are as defined in Figure~\ref{figD}.
A typical observation error bar is shown.
The observations are shown uncorrected for reddening.
The arrow on the error bar shows how
the data would shift if corrected for a reddening of $E(B-V) = 0.2$
mag. The diagonal arrow on the model curve shows how the model shifts
when the metallicity is reduced to 0.2 solar.}
\label{figDD}
\end{figure}

There is no indication that the relation between [OII] EW and the
slope $b$ in cluster galaxies differs from that in field galaxies.
This is as expected since the relationship is controlled largely by
the amount of star formation within a given galaxy and not by larger
scale environmental conditions.

\subsection{Star Formation Rates}
\label{sec:sfr}

We compute model H$\beta$ intensities and rest equivalent widths by
assuming Case B conditions (no Lyman-line photons escape the nebula
but are converted into L$\alpha$ or 2-photon plus high level hydrogen
line emission) in the ionized gas and use the ionizing flux indicated
by the BC96 models in the 230 to 912\AA\ range. It is assumed that
there is no 2-photon emission.  No reddening corrections are made.

Detailed calculations of the I([OII])/I (H$\beta$) and
I([OIII])/I(H$\beta$) ratios have been made for example by McCall,
Rybski, and Shields (1985).  McGaugh (1991) and Olofsson (1997) have
concentrated on the combined ratio (I([OII]) +
I([OIII]))/I(H$\beta$). These show that at near-solar oxygen
abundances, the I([OII])/I(H$\beta$) and (I([OII]) +
I([OIII]))/I(H$\beta$) ratios are near a maximum and there is only a
small dependence on the oxygen abundance.  A factor 10 in abundance
difference only produces a factor 1.6 in the intensity ratio.
Our observed average ratios I([OII])/I (H$\beta$) and
I([OIII])/I(H$\beta$) are best fitted with models with about twice the
solar O/H ratio. The uncertainty in this number is, however, probably
at least a factor 2.
 
Since the Bruzual and Charlot models are constructed with a known
absolute star formation rate, it is straight forward to calculate the
equilibrium star formation rate as a function of the emission line
intensities.  One finds

\begin{equation}
  SFR = I({\rm H}\beta)\times 1.49\times 10^{-41}
\end{equation}

\noindent where the intensity, $I$, is in units of erg s$^{-1}$ and
the star formation rate is in M$_{\odot}$ yr$^{-1}$.

Using the average relation between I$({\rm H}\beta)$ and I[OII] and
I$({\rm H}\beta)$ and I[OIII] from \S\ref{sec:emeqw}, one can
immediately obtain the SFR using the I[OII] or I[OIII] intensities.

\begin{equation}
  SFR = I(O[II])\times 6.80\times 10^{-42}.
\end{equation}

\begin{equation}
  SFR = I(O[III])\times 12.70\times 10^{-42}.
\end{equation}

\noindent Gallagher, Hunter, \& Bushouse (1989) use a constant of
$6.5\times10^{-42}$ for the [OII] line while Kennicutt (1992) uses a
much larger number $20\times10^{-42}$.  No allowance has been made for
foreground reddening. If the reddening corresponds to $E_{B-V}=0.24$
then the SFR increases by approximately a factor 2.2.  The SFR
estimates are given in column 17 in Tables 2, 3, and 4.  Where
possible, the [OII] line has been used; however, in low redshift cases
where [OII] is not observed, the H$\beta$ and [OIII] lines are
employed with the [OIII] result being given twice the weight of the
H$\beta$ result since the former line is on average twice as strong as
the latter while the continuum uncertainties are comparable.

\begin{figure}
\plotone{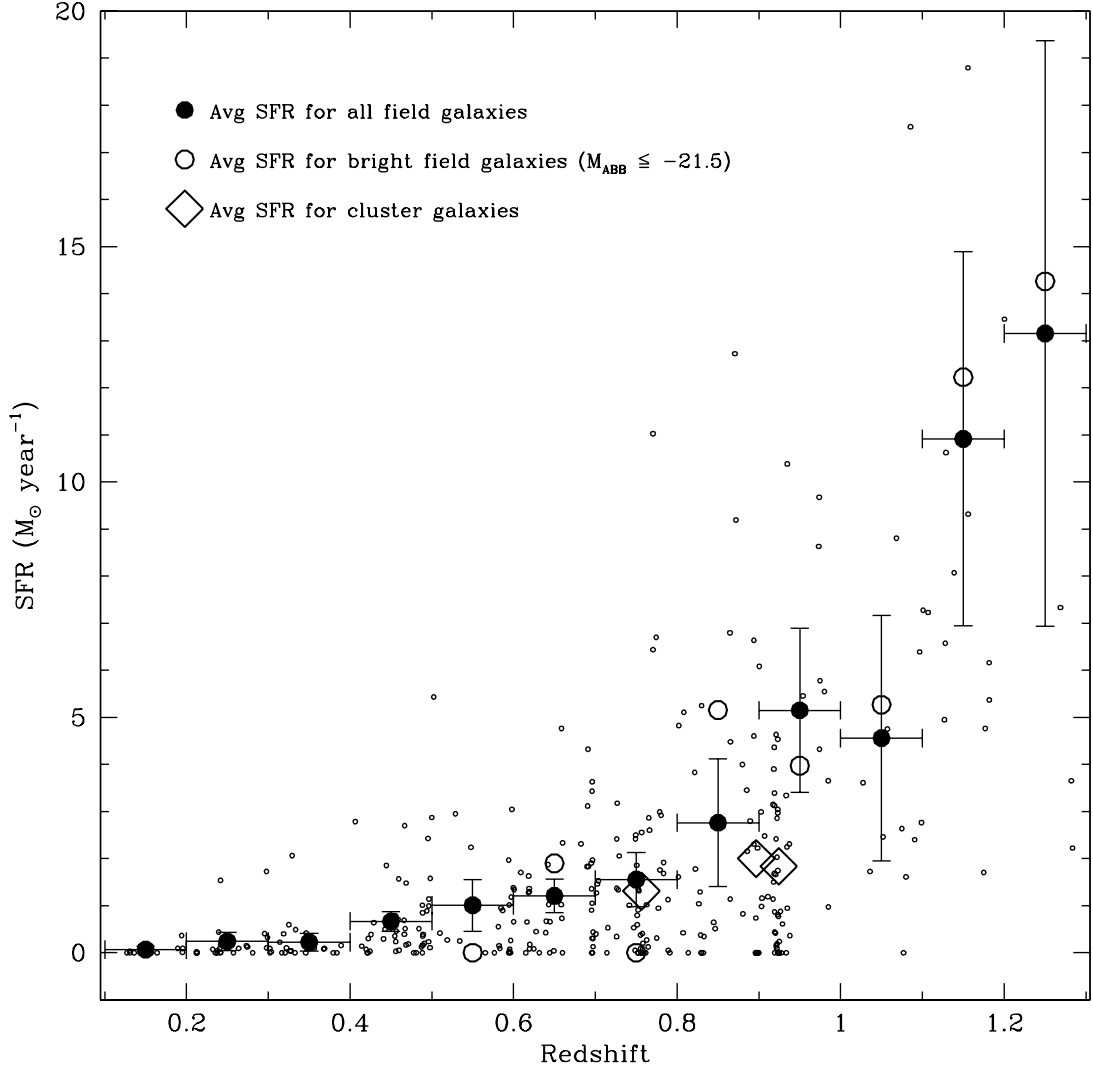}
\caption{The star formation rate, in M$_{\odot}$ yr$^{-1}$, as a
function of redshift for all galaxies in the sample.  The large solid
circles (with error bars) are the mean SFR values for field galaxies
in redshift intervals of 0.1.  The errors are the formal 2$\sigma$
uncertainties. The large diamond symbols are the mean SFR for the
galaxies in all three clusters. The large open circles are the mean
SFR values for field galaxies with $z \ge 0.7$ and $M_{ABB} <
-21.5$. The small open circles are the SFR data for each galaxy.}
\label{figsfrvsz}
\end{figure}

The SFR increases slowly with increasing redshift up to $z = 0.8$ and
exhibits a dramatic increase with redshift above $z = 0.8$. 
This behavior, shown in Figure~\ref{figsfrvsz}, 
is largely a consequence of an increasing {\it upper} envelope
to the SFR with increasing redshift coupled with the loss of intrinsically low
luminosity galaxies at higher $z$ due to the survey flux limit.
In order to make a less biased study of the SFR across a large range 
in redshift and absolute luminosity, it is necessary to normalize the SFR
value. Ideally, the best normalization would be to divide the SFR by
the galaxy mass.  Unfortunately, galaxy mass information is not
available for this sample.  The next best normalizing factor would be
the absolute K$'$--band luminosity. This would work well if the
mass-to-light ratio using the K$'$ luminosity was nearly constant. The
Bruzual and Charlot models show that this is approximately the case
only for a fairly narrow range of models in age and decay time tau.
Kelson \etal (1997) and van Dokkum \etal (1998) find that the V and I
band mass-to-light ratios of the early type cluster galaxies have
evolved passively since at least $z \sim 1.2$.

K$'$ photometry is only available for the subset of our galaxies which
lie within the WFPC2 imaging survey region. To derive K$'$ photometry
for the full sample here, it would be necessary to make extrapolations
of our BVRI data to K$'$ using the best fit stellar population models.
A preferred approach would be to use our $ABB$ photometry, which is
derived (mostly) by interpolation and, hence, constrained directly by
our observations.  We can then use the Bruzual and Charlot models and
our fits of the observations to these models to determine whether
using $ABB_{absol}$ and a normalized star formation rate defined as

\begin{equation}
	SFRN = SFR/10^{-0.4(ABB_{absol}+20)}
\end{equation}

\noindent is an acceptable surrogate for a K$'$ based normalization.
Using our constant age, variable tau model fits, we calculate the
value of $ABK'_{absol}$ using model $ABB_{absol} - ABK'_{absol}$
colors for objects in the \cldelta\ field.  We then calculate a
normalized star formation rate SFRK$'$ as defined in the equation above
but with $ABK'_{absol}$.  A linear $ABK'_{absol}$ versus $ABB_{absol}$
relation provides a very good fit to the data but the scatter is
$\pm25\%$. The SFRN -- SFRK$'$ relation is also linear with a scatter
of about $\pm15\%$.  The rare very blue galaxies with $\tau \ge 5.0$
lie significantly beyond this scatter. Had we used constant tau models
and allowed the galaxy age to vary, we still would get good linear
relations but the scatter would be somewhat larger. For statistical
purposes, therefore, it appears that using $ABB_{absol}$ to normalize
the SFR is an adequate alternative to a K$'$ based normalization.
We note that the equivalent width of the [OII] line 
is approximately proportional to SFRN: EW[OII] $\approx$ 27.6$\times$SFRN.

A further question is how nearly the luminosity calculated from the
$K'$ magnitude is proportional to the mass.  Again, using the Bruzual
and Charlot models, it is possible to investigate this for the range
of models and ages which are relevant.  At a given redshift it is
found that it does represent the mass with a scatter of $\pm20-30\%$.
It does however change systematically with redshift.  Between $z = 0.50$
and $z = 1.30$ $ABK'_{absol}$ brightens by 0.53 mag and is linear with $z$.
This is because in our constant age models the galaxies get younger as
redshift increases. In a constant tau scenario this effect is not present
because there is a range of ages at each redshift.

\begin{figure}
\plotone{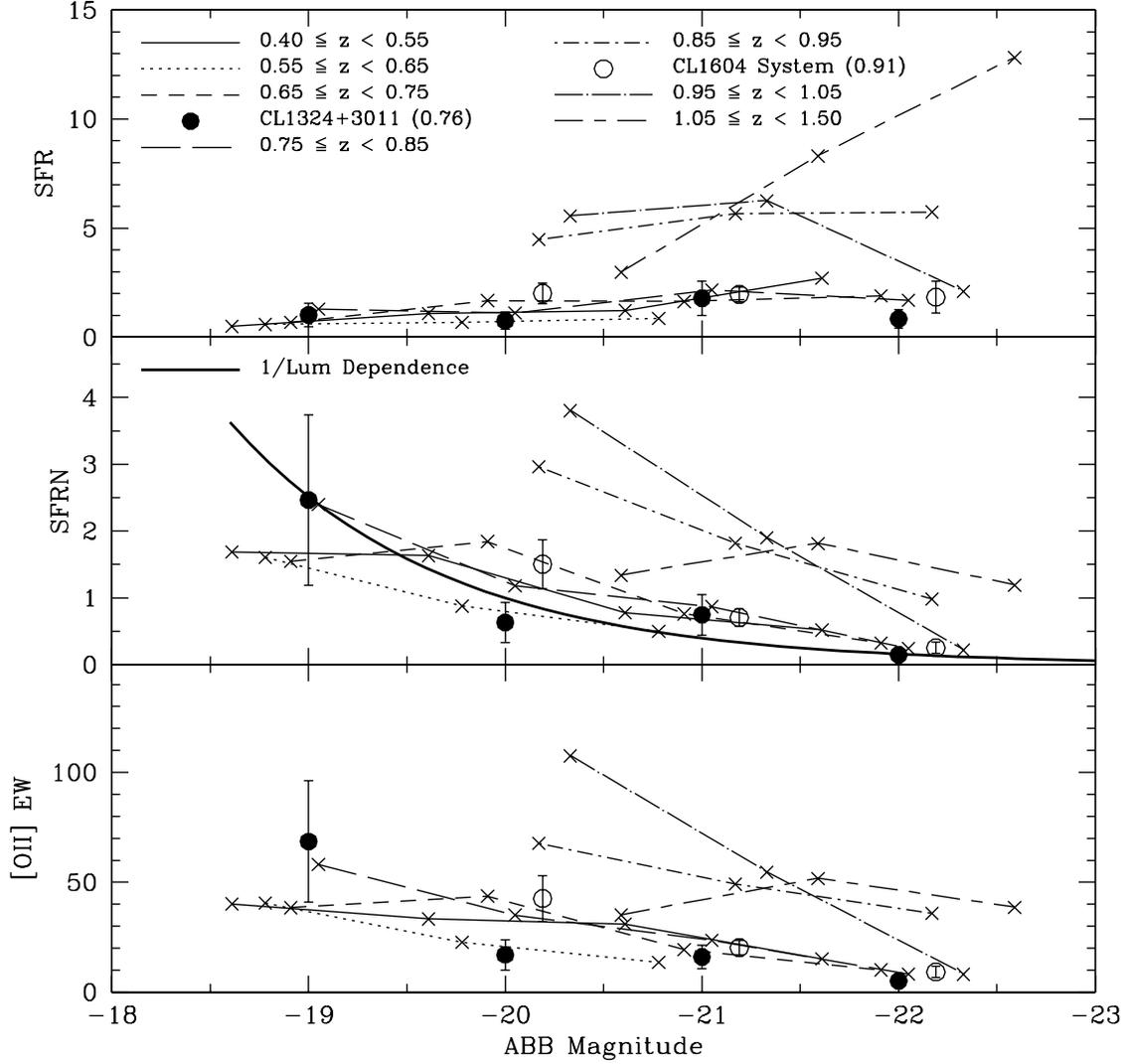}
\caption{Various star formation indicators as a function of ABB magnitude and
redshift. The upper plot shows the direct star formation rate (M$_{\odot}$ yr$^{-1}$);
the middle plot shows the normalized star formation rate, SFRN in units of M$_{\odot}$
yr$^{-1}$ per unit ABB luminosity, and the lower plot shows the absolute value
of the [OII] line equivalent width (in \AA).
Data for field galaxies are plotted as lines, data for the
cluster members are shown as points with 1$\sigma$ errors
(the CL1604 system results include data for \cldelta\ and \cleps).
In the SFRN vs ABB plot, the heavy solid line is the trend expected for a fixed
star formation rate. }
\label{figsfrn_vs_abb}
\end{figure}

To investigate how the star formation rate, the normalized star formation
rate (SFRN), and the [OII] line equivalent width change with galaxy luminosity 
and redshift, we have chosen to generate averages in bins corresponding to
approximately 0.1 in $z$ and 1 magnitude in luminosity.  The luminosity
bins allow for the expected evolution in luminosity with redshift.
The SFRN results are listed in Table 10 and all three SFR indicators are plotted in
Figure~\ref{figsfrn_vs_abb}.  There are two general trends: (1) at
a given luminosity, both the [OII] EW and the SFRN increase as $z$ increases; 
(2) at a given redshift, both the [OII] EW and the SFRN decrease 
as the luminosity increases. The remarkable aspect of this latter trend is
the uniformity of the nearly linear decline in SFRN with increasing ABB luminosity 
for galaxies with $z \simless 0.9$. This result implies that, on average,
the SFR is effectively independent of galaxy luminosity over the range
$-18 \ge {\rm ABB} \ge -22$. This is explicitly demonstrated in the upper
plot of Figure~\ref{figsfrn_vs_abb}. Sullivan \etal (2000) find a similarly
weak dependence of [OII] equivalent width on galaxy luminosity at $z < 0.2$.

The increase in SFRN with redshift, at a fixed ABB luminosity, is consistent with 
the predictions of our BC model fits. Naturally, their are
important differences between galaxies with different spectral
and photometric characteristics. For example, 
one expects the SFRN to be dependent on the color of the galaxy,
being larger for galaxies which are bluer.  Since the equivalent width
of [OII] is approximately proportional to SFRN, the result is very much like that in
Figure~\ref{figDD}.  In the clusters in particular, for slope $b \ge
11$, the average SFRN is 0.20 while for $b < 5$, every object has SFRN
$> 1.0$.  Red objects have very little star formation while in blue
objects the star formation rate is high.  One obtains a similar result
from the spectral classifications. The average values of SFRN are
1.56, 0.79, 0.61, and 0.25, respectively, for spectral classes {\it
``a"}, {\it ``a+k"},{\it ``k+a"}, and{\it ``k"}.

\section{Cluster Structure}
\label{sec:clstruc}

\begin{figure}
\plotone{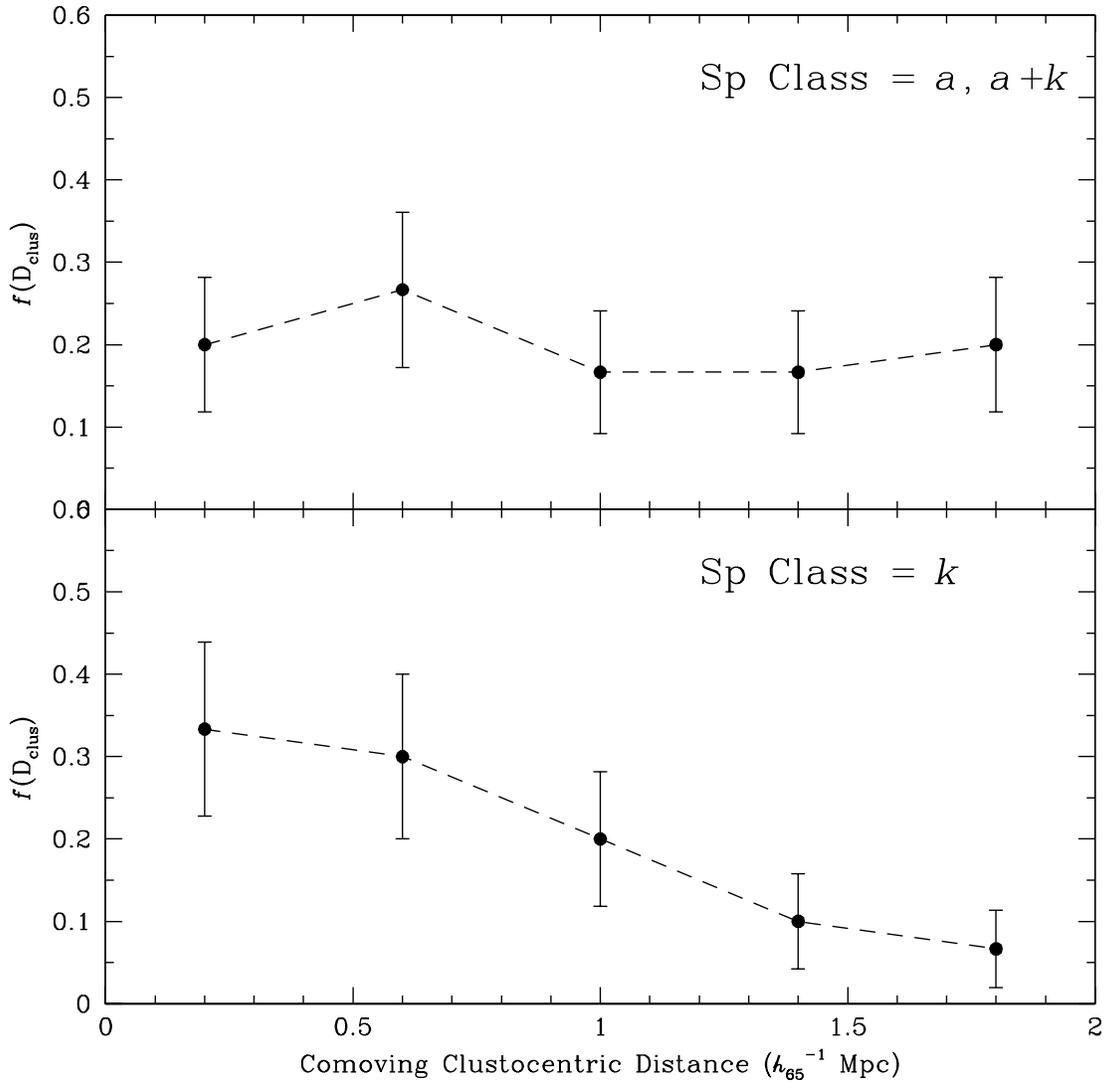}
\caption{The distribution of galaxy spectral type as a function of the comoving
clustocentric radius. Data for all three clusters are combined in this figure.}
\label{figsrad}
\end{figure}

Cluster formation is believed to involve both a significant amount of
infall of the surrounding field galaxies into the cluster's potential 
as well as environmental processing of the cluster members as they undergo
ram pressure stripping, tidal disruption, and merging. 
This hypothesis predicts that there should be 
correlations between the spectral properties of the cluster members
and their clustocentric distance. The amplitude of the correlations tell
us something about the fraction of cluster members which have been
recently accreted and/or the degree to which environmental processes
alter the galaxy characteristics. 
Figure~\ref{figsrad} shows the distribution of
galaxy spectral class as a function of the comoving clustocentric
distance. Galaxies of spectral class {\it ``k"} appear to be somewhat
more centrally concentrated than the bluer galaxies of spectral
classes {\it ``a"} and {\it ``a$+$k"}. The difference, however, is not
significant in our current sample. A KS test yields a 42\% probability
that galaxies with spectral class {\it ``k"} are drawn from the same
spatial distribution as those with spectral classes {\it ``a"} or {\it
``a$+$k"}. However, the factor of $\sim 3$ decline in the fraction 
of {\it ``k"}-type galaxies as the clustocentric radius increases 
from 250\ih65 kpc to 1.5\ih65 Mpc is significant.

\begin{figure}
\plotone{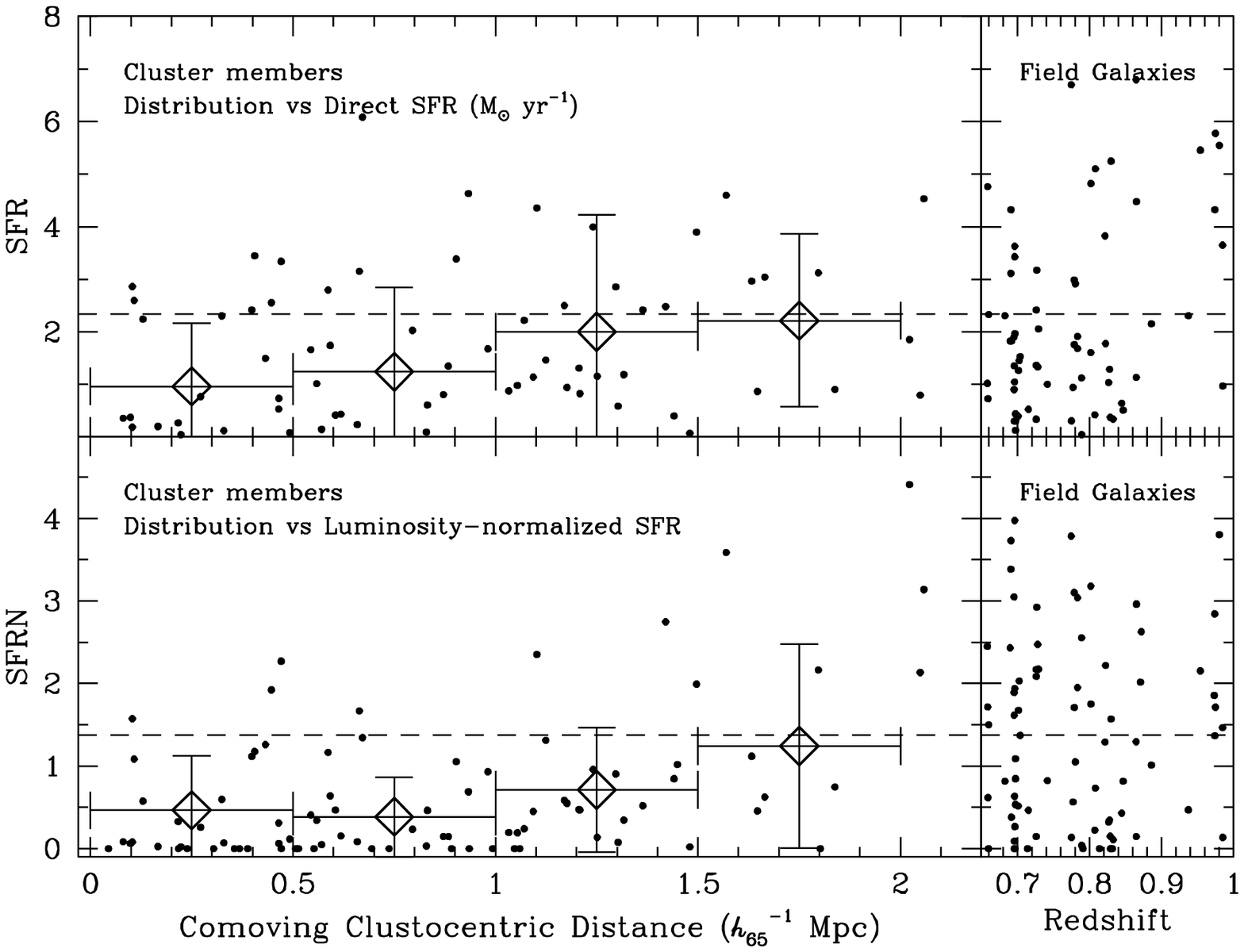}
\caption{The direct star formation rate, SFR (upper plot),
and the normalized star formation rate, SFRN (lower plot),
as functions of the comoving clustocentric radius. The large
diamonds show the mean values in four (500\ih65 kpc) radial bins.
The trends for both these measures of star formation are similar.
Vertical errors on the large diamonds indicate the standard deviation about the
mean, horizontal errors show the radial bin size.
The small points are the data for individual objects. Also shown, at right,
are the SFR and SFRN as functions of redshift for field galaxies at redshifts similar
to those of the clusters ($0.65 \le z \le 1$).
The dashed lines are the mean SFR and SFRN values for the field galaxies.}
\label{figsfrn}
\end{figure}

We can also look at the values of the SFR and SFRN as functions of the projected distance 
of a galaxy from the cluster center. The data are shown in Figure~\ref{figsfrn}.
The data for the three clusters are shown
on the left side of the figure. On the right side, for comparison, 
are the SFR and SFRN values for field galaxies in the redshift 
range $0.65 \le z \le 1.00$. 
For comoving clustocentric radii less than 1\ih65 Mpc, a large
majority of the cluster objects have low SFRN relative to their field
galaxy counterparts.  At large distances there are very few low SFRN
objects. The few high SFRN objects projected near the center of the
cluster could be outliers in the distribution, recently infalling
galaxies, or foreground and background objects, although their
redshifts are consistent with their being cluster members. A
comparison of the cluster galaxies at large distances and field
galaxies shows them to be quite similar, although the cluster numbers
are quite small.  Similar results have been produced for clusters and
field galaxies with redshifts from 0.18 to 0.55 by Balogh \etal
(1997), Morris \etal (1998), Balogh \etal (1998), and Balogh \etal
(1999). They find the same suppression of the SFR in clusters relative
to the field at the same redshifts. They suggest that this is caused
by ram pressure or tidal stripping of the gas from galaxies as they
first fall into the cluster potential or pass near the cluster center.
Balogh \etal (1998) find that even in the outer regions of the
clusters the SFR is less than in the field whereas our data suggest
that they are much the same.  This may be an evolutionary effect in
which the process of suppression of the SFR in clusters has proceeded
further at $z = 0.5$ than at the much earlier time corresponding to
$z = 0.9$. Larger surveys, sampling systems with a broad range in ICM
properties, are needed to fully understand the nature of the observed
trends.

\section{Conclusions}
\label{sec:conclude}

Extensive data from a joint HST/Keck imaging and spectroscopic survey of
9 distant clusters has provided a wealth of observational constraints
on galaxy and cluster evolution. In this paper, we have focused on the 
properties of the galaxies in and around the 3 most massive clusters in
our survey: \clalpha\ ($z = 0.757$), \cldelta\ ($z = 0.897$), and 
\cleps\ ($z = 0.924$). Our most important results are summarized below.
 
\begin{itemize}
\item The central velocity dispersions of the clusters are accurately
measured from the redshifts of $\sim$20 to 40 member galaxies. The
dispersions lie in the range 900 -- 1300 \kms and the resulting
kinematic mass estimates are $\simgreat 5 \times 10^{14}$\ih65
M$_{\odot}$ (see Table 5 and Figure~\ref{figvdisp}).  The 
bolometric x-ray luminosities of these clusters are: 
log$(L_{x,Bol}) = 43.95$ for \clalpha,
log$(L_{x,Bol}) = 44.05$ for \cldelta, and
log$(L_{x,Bol}) \le 43.90$ for \cleps 
(the uncertainties in the logarithm of the bolometric luminosity are about $\pm0.5$).
These values are low for the derived velocity dispersions of these
clusters.  For example, the low-$z$ $\sigma - {\rm T}_x$ relation (Donahue \etal
1998) would predict our clusters should have kinetic ICM temperatures
in the range 5 -- 10 keV, suggesting bolometric x-ray luminosities in excess of
$10^{45}$ erg s$^{-1}$ (based on the ${\rm L}_x - {\rm T}_x$ relation
from Mushotsky \& Scharf 1997). Specifically, the predicted bolometric luminosities 
based on the Mushotsky \& Scharf results
\footnote{The values of the predicted bolometric luminosity are derived
by taking the mean of the least square fits using $L_{x,Bol}$ and then 
$T_x$ as the independent variable.} 
are log$(L_{x,Bol})$ = 45.0, 45.4, and 44.7 for 
\clalpha, \cldelta, and \cleps, respectively -- about an order of magnitude
larger than what is actually observed (the factor can lie in the range
3 to 60, depending on the cluster, due to the error in the observed quantity).
However, this is not uncommon in optically selected clusters at $z > 0.7$
(\eg see Holden \etal 1997). Possible
explanations for this may include an enhanced population of 
optically rich, x-ray faint clusters at $z > 0.6$, more efficient detection
of galaxy-rich clusters in which the dynamical state of the ICM 
is still evolving when selecting clusters in optical passbands, 
and a higher fraction of clusters at $z \sim 0.8$ in which the 
infall of field galaxies is still a significant process.

\item The cluster M/L ratios are consistent with those derived for
their low-$z$ counterparts, lying comfortably between 100 -- 500 (see
Table 6).  The systematic errors associated with determining these
ratios prevent any conclusive statements about evolution of the M/L
ratios. However, large M/L ($>1300$) ratios, expected if $\Omega_m =
1$ on cluster scales, are strongly ($>99$\% C.L.) rejected.

\item The rest-frame B-band characteristic magnitude of the integrated
field galaxy luminosity function evolves with redshift roughly as $M^*(z) =
M^*(0) - \beta z$ where $1 < \beta < 1.5$ (see Table 7 and
Figures~\ref{figlffit} and~\ref{figlfevol}). Caution is recommended in
the interpretation of this observation as we do not have a sufficient
number of objects to subdivide our analysis by galaxy color or
spectral class. It is well known that the local LF is a strong
function of color (\eg Blanton \etal 2001) and evolution of the LF has
already been shown to be color dependent (\eg Lilly \etal 1995). We
note, however, the our observations of the brightest cluster galaxies
show that their rest-frame B-band luminosity evolves with redshift in
a very similar fashion. The tau0.2 model predicts an increase in BCG
luminosity which is nearly identical to the simple LF evolution with
$\beta = 1$; the tau1.0 model for BCG evolution is nearly identical to
the $\beta = 1.5$ scenario. The observed evolution of the absolute BCG
magnitudes is best fit by solar metallicity models 
with $0.2 \simless \tau < 0.6$. Our data also suggest that 
the $M^*_{ABB}$ in clusters has dimmed
by about 0.7 mag between $z \sim 0.8$ and the current epoch. 

\item The mean spectral characteristics of the cluster galaxies, such
as the metal-line equivalent widths, $J_u$ jump parameter, and derived
(normalized) star formation rates, are well correlated with their
broadband photometric properties, such as the slope $b$ value (see
Table 8 and Figures~\ref{figsptypevsb},~\ref{figD},~\ref{figjujl},
and~\ref{figDD}). These correlations are a straight forward
consequence of the star formation process -- galaxies with active star
forming regions have spectra with oxygen line emission and Balmer
absorption. These same galaxies have blue SEDs as a consequence of
their young stellar population. Similarly, the light from older
galaxies, which typically have strong metal-line absorption features,
is dominated by a red stellar population. These correlations are also
observed in field galaxies.

\item The spectroscopic and photometric properties of the cluster
galaxies are well fit by the Bruzual-Charlot solar metallicity,
constant-age (4.8 Gyr at $z = 0.9$), variable tau models. Models with
sub-solar metallicity are not strongly rejected so long as long as the
metallicity is at least 0.2 of solar. A significant amount of dust
extinction at $z \sim 0.9$ is not likely -- the observations can be
fit acceptably by the solar metallicity models so long as E$(B-V)
\simless 0.2$ mag.

\item All the clusters host a significant early-type galaxy
population, although an envelope of red elliptical-like galaxies, so
prominent in many $z \le 0.5$ clusters, is only weakly detected at
these redshifts (see Figures~\ref{figCC} and~\ref{figS}). Indeed, many
of the spectroscopically confirmed cluster members are blue (slope $b\
< 7$, $V-R < 0.5$) and some remarkably so ($V-R \approx 0$).  These
observations are consistent with the predictions from coeval tau
models: the intrinsic scatter in the logarithmic slope of the
(optical) spectral energy distributions of the redder cluster members is
expected to decrease by a factor $\sim 2$ between $z = 0.9$ and $z = 0.5$
and by a further factor of $\sim 2$ between $z = 0.5$ and the current
epoch. We note that the contrast of the red elliptical sequence is
enhanced significantly if a NIR passband is combined with an optical
passband in the definition of the CM diagram (\eg Gladders \& Yee
2000).

\item The average I([OII])/I(H$\beta$) ratio is very similar to that
found in nearby galaxies while the average I([OIII])/I(H$\beta$) is
somewhat higher.  These ratios are best fit by models with a
metallicity of twice solar, although the uncertainty is at least a
factor 2.

\item A star formation rate, SFRN, normalized to the galaxy's
luminosity at the rest $B$ wavelength, is found to increase as the
redshift increases and decrease as the luminosity increases
(see Table 10 and Figure~\ref{figsfrn_vs_abb}).
In addition, models show that, on average, the SFRN is predicted
to be roughly proportional to the SFR per unit mass our sample of galaxies.
The SFR and SFRN are correlated with galaxy color and spectral classification -
redder galaxies exhibit, on average, weaker star formation activity and stronger
metal line absorption.
One remarkable aspect of these results is the uniformity of the decline of SFRN
with luminosity over a large range of ABB magnitude
and redshift. This implies that the average SFR per galaxy in this sample is
nearly independent of galaxy redshift and luminosity in the ranges
$0.4 \simless z \simless 0.9$ and $-18 \ge {\rm ABB} \ge -22$.

\item For field galaxies with $0.40 < z \le 0.85$, 65\% are
active ([OII] equivalent width greater than 15\AA). 
For field galaxies with $z > 0.85$, the fraction is 79\%.
Within the central 1.5 \ih65 Mpc regions of our three distant
clusters, the fraction of active galaxies is 45\%. This is
substantially higher than the 10 to 20\% active galaxy component seen
in the centers of $0.2 < z < 0.55$ clusters but is lower than that seen
in the field at redshifts comparable with those of our distant clusters.  
Within the central comoving cluster distance of 1.0\ih65 Mpc a large fraction of the
cluster galaxies have low values of SFRN (relative to field galaxies
at similar redshifts) and have spectra which 
exhibit strong CaII K, $\lambda$3835 and g-band absorption but
exhibit little or no H$\gamma$ absorption. Above 1.0\ih65 Mpc, on the other
hand, the average SFRN in cluster galaxies 
is nearly as high as that in field galaxies in the same
redshift range. The smooth melding of the star forming properties of
galaxies in the outskirts of the clusters with that in galaxies in the general
field (see Figure~\ref{figsfrn}) coupled with the lower overall fraction of
active galaxies in the cluster cores 
can be understood if a) the accretion of field galaxies by the clusters 
is a significant process at these intermediate redshifts and/or b)
if the volume containing the majority of gas-poor cluster members (due to
environmental processes such as ram pressure stripping) evolves with time.
In the latter case, we would conclude that at $z \sim 0.8$ the effects of
cluster induced gas stripping have been largely confined to the central
1\ih65 Mpc region whereas at $z < 0.5$ the effects of these processes 
encompass a significantly larger volume about the cluster center
(\eg Balogh \etal 1998).

\item  We detect a factor of $\sim 3$ 
decline in the fraction of {\it ``k"}-type galaxies
as the clustocentric radius increases from 250\ih65 kpc to 1.5\ih65 Mpc
(see Figure~\ref{figsrad}).
This is further evidence in support of the importance of environmental
effects in cluster cores on star formation activity in galaxies. However,
a similar trend would be expected in a hierarchical structure formation 
scenario in which the most massive galaxies form first and are thus the oldest
(and, hence, most centrally located) members of the cluster.

\end{itemize}

\acknowledgments

LML is supported by NASA through Hubble Fellowship grant
HF-01095.01-97A from the Space Telescope Science Institute, which is
operated by the Association of Universities for Research in Astronomy,
Inc., under NASA contract NAS 5-26555.

Observational material for this paper was obtained at the W. M. Keck
Observatory, which is operated as a scientific partnership between the
California Institute of Technology, the University of California, and
the National Aeronautics and Space Administration.  It was made
possible by the generous financial support of the W. M. Keck
Foundation.

\clearpage 

\begin{table}
\begin{center}
Table 1. Summary of Spectroscopic Observations
\end{center}
\begin{center}


\clearpage

\hoffset 0.0in
\begin{table}
\begin{center}
Table 5. Cluster Dynamical Parameters
\begin{tabular}[h]{cccccccccrrrccc}
\hline
\hline
 & & & & $M_{PW}$ & $M_{PM}$ & $M_{RW}$ & Radius \\
Cluster & N$_z$ & $\overline z$ & $\sigma$ & \multicolumn{3}{c}{($10^{14}$\ih65 
M$_{\odot})$} & (\ih65 kpc) \\ 
\hline
CL1324+3011 & 16 & 0.7568&1037$^{+258}_{-150}$ & 5.89$^{+2.94}_{-1.71}$ & 
7.48$\pm0.58$ & 11.8$\pm0.92$ & 385 \\
CL1324+3011 & 24 & 0.7561& 942$^{+178}_{-115}$ & 8.15$^{+3.11}_{-2.03}$ & 
9.71$\pm0.49$ & 13.8$\pm0.69$ & 770 \\
CL1324+3011 & 32 & 0.7565&1058$^{+166}_{-114}$ & 13.9$^{+4.37}_{-3.02}$ & 
31.1$\pm2.49$ & 28.0$\pm2.25$ & Unlimited\\
 & & & & & & \\
CL1604+4304 & 11 & 0.8964& 921$^{+303}_{-155}$ & 3.77$^{+2.58}_{-1.46}$ & 
6.48$\pm1.17$ & 14.2$\pm2.57$ & 385 \\
CL1604+4304 & 19 & 0.8967&1300$^{+286}_{-173}$ & 10.6$^{+4.78}_{-3.03}$ & 
42.0$\pm4.52$ & 48.2$\pm5.18$ & 770 \\
CL1604+4304 & 22 & 0.8967&1226$^{+245}_{-154}$ & 12.0$^{+4.91}_{-3.23}$ & 
38.9$\pm3.58$ & 48.2$\pm4.43$ & Unlimited \\
 & & & & & & \\
CL1604+4321 & 17 & 0.9246& 994$^{+236}_{-140}$ & 5.08$^{+2.54}_{-1.62}$ & 
6.68$\pm0.72$ & 11.6$\pm1.25$ & 385 \\
CL1604+4321 & 35 & 0.9243& 922$^{+137}_{-96} $ & 9.31$^{+2.80}_{-2.00}$ & 
13.9$\pm0.66$ & 21.5$\pm1.03$ & 770 \\
CL1604+4321 & 41 & 0.9243& 935$^{+126}_{-91} $ & 11.8$^{+3.22}_{-2.34}$ & 
20.9$\pm1.35$ & 24.9$\pm1.62$ & Unlimited\\
\hline
\end{tabular}
\end{center}
\end{table}

\begin{table}
\begin{center}
Table 6. Cluster Mass-to-Light Ratios 
\begin{tabular}[h]{cclll}
\hline
\hline
        &Radius & $M_{PM}/L_B$ & $M_{PM}/L_V$ & $M_{PM}/L_R$ \\   
Cluster & (\ih65 kpc) & \multicolumn{3}{c}{(\h65 M$_{\odot}$/L$_{\odot}$)}\\
\hline
CL1324+3011 & 385 & $347\pm49 $ & $327\pm85 $ & $527\pm106$ \\
            & 770 & $285\pm40 $ & $255\pm66 $ & $416\pm84 $ \\
CL1604+4304 & 385 & $208\pm90 $ & $144\pm50 $ & $159\pm54 $ \\
            & 770 & $127\pm28 $ & $134\pm24 $ & $319\pm58 $ \\
CL1604+4321 & 385 & $248\pm45 $ & $168\pm47 $ & $500\pm161$ \\
            & 770 & $ 96\pm8  $ & $103\pm14 $ & $365\pm57 $ \\
\hline
\end{tabular}
\end{center}
\end{table}

\begin{table}
\begin{center}
Table 7. Galaxy Luminosity Function Parameters (with $\alpha = -1.15$)
\begin{tabular}[h]{ccccccc}
\hline
\hline
 & & & $M_{ABB}^*$ & $\phi^*$ &  & \\
$z$ range &$\overline z$ & N$_{\rm gal}$ & (+5log(\h65)) & (\ch65 Mpc$^{-3}$) & 
$\chi_{\nu}^2$ & $\left< V / V_{max} \right>$ \\
\hline
$0.1 \le z < 0.5$   & 0.362 & 103 & $-20.35\pm0.24$ & 0.024 & 1.42 & 0.27\\
$0.5 \le z < 0.7$   & 0.630 &  69 & $-20.80\pm0.21$ & 0.017 & 1.10 & 0.35\\
$0.7 \le z \le 1.0$ & 0.820 &  57 & $-21.15\pm0.23$ & 0.011 & 0.19 & 0.49\\
Cluster Members     & 0.859 &  91 & $-21.44\pm0.17$ & \nodata & 2.83 & 0.43\\
\hline
\end{tabular}
\tablecomments{Results for $0.7 \le z \le 1.0$ bin exclude the cluster members.}
\end{center}
\end{table}

\clearpage
\begin{table}
\begin{center}
Table 8. Mean Spectral Properties of Distant Cluster Galaxies
\begin{tabular}[h]{l|c|ccc|cc|c}
\hline
\hline
 & & EW $\lambda3835$ &
EW CaII K &
EW g-band &
$J_l$ &
$J_u$ & \\
Subset & $\overline b$ & \multicolumn{3}{c|}{(Angstroms)} & 
\multicolumn{2}{c|}{(mag)} & ABB \\
\hline
Blue ($b \le 10$) & ~6.06  & 2.41   &  3.01  & 1.89   &  0.43   & 0.20   & -20.83 \\
   & ($\pm$0.36) & ($\pm$0.41) & ($\pm$0.52) & ($\pm$0.57) &  ($\pm$0.03) & ($\pm$0.02) & ($\pm$0.11) \\
Red  ($b \ge 11$) & 13.93  & 5.37   &  7.64  & 3.62   &  0.47   & 0.43   & -21.53 \\
   & ($\pm$0.31) & ($\pm$0.48) & ($\pm$0.30) & ($\pm$0.41) &  ($\pm$0.02) & ($\pm$0.01) & ($\pm$0.11) \\
\hline
Type ``{\it a}"   & ~4.96  & 1.88   & 1.92   & 1.69   &  0.39   & 0.15   & -20.82 \\
   & ($\pm$0.57) & ($\pm$0.82) & ($\pm$0.93) & ($\pm$1.47) &  ($\pm$0.04) & ($\pm$0.02) & ($\pm$0.21) \\
Type ``{\it a+k}" & ~7.24  & 2.84   & 1.99   & 0.81   &  0.51   & 0.20   & -20.93 \\
   & ($\pm$0.79) & ($\pm$0.57) & ($\pm$0.48) & ($\pm$0.34) &  ($\pm$0.07) & ($\pm$0.03) & ($\pm$0.17) \\
Type ``{\it k+a}" & 10.22  & 3.55   & 5.14   & 2.04   &  0.50   & 0.30   & -21.31 \\
   & ($\pm$0.93) & ($\pm$0.57) & ($\pm$0.40) & ($\pm$0.61) &  ($\pm$0.03) & ($\pm$0.02) & ($\pm$0.16) \\
Type ``{\it k}"   & 12.89  & 5.50   & 8.19   & 4.40   &  0.46   & 0.44   & -21.50 \\
   & ($\pm$0.53) & ($\pm$0.53) & ($\pm$0.30) & ($\pm$0.37) &  ($\pm$0.02) & ($\pm$0.01) & ($\pm$0.11) \\
\hline
\end{tabular}
\end{center}
\end{table}

\begin{table}
\begin{center}
Table 9. Predicted Evolution in BCG Luminosity: $M_{ABB}\ vs\ z$
\end{center}
\begin{center}
\begin{tabular}[h]{c|ccc|ccc}
\hline
\hline
    &\multicolumn{3}{|c|}{$H_o = 65$} &\multicolumn{3}{c}{$H_o = 80$} \\
$z$ & tau0.2  & tau0.6  & tau1.0 & tau0.2  & tau0.6  & tau1.0 \\
\hline
0.033 & $-21.59$ & $-21.59$ & $-21.59$ & $-21.59$ & $-21.59$ & $-21.59$ \\
0.190 & $-21.79$ & $-21.81$ & $-21.83$ & $-21.79$ & $-21.81$ & $-21.86$ \\
0.320 & $-21.95$ & $-21.97$ & $-22.01$ & $-21.94$ & $-22.00$ & $-22.08$ \\
0.490 & $-22.12$ & $-22.17$ & $-22.26$ & $-22.13$ & $-22.23$ & $-22.40$ \\
0.620 & $-22.25$ & $-22.31$ & $-22.46$ & $-22.26$ & $-22.40$ & $-22.65$ \\
0.760 & $-22.37$ & $-22.47$ & $-22.68$ & $-22.41$ & $-22.61$ & $-22.91$ \\
0.840 & $-22.45$ & $-22.56$ & $-22.81$ & $-22.50$ & $-22.72$ & $-23.06$ \\
0.900 & $-22.51$ & $-22.64$ & $-22.91$ & $-22.56$ & $-22.81$ & $-23.16$ \\
1.100 & $-22.69$ & $-22.90$ & $-23.24$ & $-22.73$ & $-23.15$ & $-23.50$ \\
1.300 & $-22.90$ & $-23.24$ & $-23.61$ & $-22.83$ & $-23.39$ & $-23.70$ \\
1.500 & $-22.98$ & $-23.41$ & $-23.78$ & $-23.05$ & $-23.69$ & $-23.94$ \\
\hline
\end{tabular}
\end{center}
\end{table}

\clearpage
\hoffset 0.0in
\begin{table}
\begin{center}
Table 10. Mean Normalized Star Formation Rate vs Redshift (M$_\odot$ yr$^{-1}$ 
L$_{ABB}^{-1}$)
\end{center}
\begin{center}
\begin{tabular}[h]{c|cccc|cccc}
\hline
\hline
$z$ range   &\multicolumn{4}{c|}{Luminosity range} &\multicolumn{4}{c}{SFRN(No. 
objects)}\\
	    &   1     &    2   &  3     &  4     &  1       & 2        &  3      
&  4   \\
\hline
0.40-0.55   &  -18.11 & -19.11 & -20.11 & -21.11 & 1.69 (23) & 1.63 (21) & 0.78 
(5) & 0.52 (1) \\
	    &  -19.10 & -20.10 & -21.10 & -22.10 & & & &	\\
0.55-0.65   &  -18.28 & -19.28 & -20.28 & -21.28 & 1.61 (9) & 0.88 (18) & 0.50 
(11) & \nodata \\
	    &  -19.27 & -20.27 & -21.27 & -22.27 & & & &	\\
0.65-0.75   &  -18.41 & -19.41 & -20.41 & -21.41 & 1.55 (6) & 1.84 (19) & 0.76 
(9) & 0.32 (2) \\
	    &  -19.40 & -20.40 & -21.40 & -22.40 & & & &	\\
\hline
CL1324      &  -18.50 & -19.50 & -20.50 & -21.50 & 2.46 (3) & 0.63 (7) & 0.75 
(16) & 0.15 (6) \\
$z = 0.76$  &  -19.49 & -20.49 & -21.49 & -22.49 & & & &	\\
\hline
0.75-0.85   &  -18.55 & -19.55 & -20.55 & -21.55 & 2.40 (4) & 1.18 (8) & 0.87 
(13) & 0.24 (3) \\
	    &  -19.54 & -20.54 & -21.54 & -22.54 & & & &	\\
0.85-0.95   &  -18.67 & -19.67 & -20.67 & -21.67 & \nodata  & 2.96 (1) & 1.82 
(2) & 0.98 (4) \\
	    &  -19.66 & -20.66 & -21.66 & -22.66 & & & &	\\
\hline
CL1604      &  -18.69 & -19.69 & -20.69 & -21.69 &\nodata & 1.50 (12) & 0.70 
(31) & 0.25 (15) \\
$z \approx 0.91$ &  -19.68 & -20.68 & -21.68 & -22.68 & & & &	\\
\hline
0.95-1.05   &  -18.83 & -19.83 & -20.83 & -21.83 & \nodata & 3.80 (1) & 1.90 (6) 
& 0.22 (3) \\
	    &  -19.82 & -20.82 & -21.82 & -22.82 & & & &	\\
1.05-1.50   &  -19.09 & -20.09 & -21.09 & -22.09 & \nodata & 1.34 (3) & 1.81 
(15) & 1.19 (14) \\
	    &  -20.08 & -21.08 & -22.08 & -23.08 & & & &	\\
\hline
\end{tabular}
\end{center}
\end{table}

\end{document}